\documentclass[amsmath,amssymb,aps,twocolumn,prd,superscriptaddress]{revtex4-2}
\usepackage[T5,T1]{fontenc}
\usepackage{graphicx} % Include figure files
\graphicspath{{Images/}}
\usepackage{dcolumn} % Align table columns on decimal point
\usepackage{bm} % Bold math
\usepackage{multirow}
\usepackage{gensymb}
\usepackage{caption}
\usepackage{subcaption}
\usepackage{csquotes}
\MakeOuterQuote{"}
\usepackage{placeins}
\usepackage{xcolor}
\usepackage{hyperref}
\hypersetup{colorlinks=true,allcolors=blue}

\begin{document}

\title{Limits on GeV-scale WIMP Annihilation in Dwarf Spheroidals with IceCube DeepCore}

\affiliation{III. Physikalisches Institut, RWTH Aachen University, D-52056 Aachen, Germany}
\affiliation{Department of Physics, University of Adelaide, Adelaide, 5005, Australia}
\affiliation{Dept. of Physics and Astronomy, University of Alaska Anchorage, 3211 Providence Dr., Anchorage, AK 99508, USA}
\affiliation{School of Physics and Center for Relativistic Astrophysics, Georgia Institute of Technology, Atlanta, GA 30332, USA}
\affiliation{Dept. of Physics, Southern University, Baton Rouge, LA 70813, USA}
\affiliation{Dept. of Physics, University of California, Berkeley, CA 94720, USA}
\affiliation{Lawrence Berkeley National Laboratory, Berkeley, CA 94720, USA}
\affiliation{Institut f{\"u}r Physik, Humboldt-Universit{\"a}t zu Berlin, D-12489 Berlin, Germany}
\affiliation{Fakult{\"a}t f{\"u}r Physik {\&} Astronomie, Ruhr-Universit{\"a}t Bochum, D-44780 Bochum, Germany}
\affiliation{Universit{\'e} Libre de Bruxelles, Science Faculty CP230, B-1050 Brussels, Belgium}
\affiliation{Vrije Universiteit Brussel (VUB), Dienst ELEM, B-1050 Brussels, Belgium}
\affiliation{Dept. of Physics, Simon Fraser University, Burnaby, BC V5A 1S6, Canada}
\affiliation{Department of Physics and Laboratory for Particle Physics and Cosmology, Harvard University, Cambridge, MA 02138, USA}
\affiliation{Dept. of Physics, Massachusetts Institute of Technology, Cambridge, MA 02139, USA}
\affiliation{Dept. of Physics and The International Center for Hadron Astrophysics, Chiba University, Chiba 263-8522, Japan}
\affiliation{Department of Physics, Loyola University Chicago, Chicago, IL 60660, USA}
\affiliation{Dept. of Physics and Astronomy, University of Canterbury, Private Bag 4800, Christchurch, New Zealand}
\affiliation{Dept. of Physics, University of Maryland, College Park, MD 20742, USA}
\affiliation{Dept. of Astronomy, Ohio State University, Columbus, OH 43210, USA}
\affiliation{Dept. of Physics and Center for Cosmology and Astro-Particle Physics, Ohio State University, Columbus, OH 43210, USA}
\affiliation{Niels Bohr Institute, University of Copenhagen, DK-2100 Copenhagen, Denmark}
\affiliation{Dept. of Physics, TU Dortmund University, D-44221 Dortmund, Germany}
\affiliation{Dept. of Physics and Astronomy, Michigan State University, East Lansing, MI 48824, USA}
\affiliation{Dept. of Physics, University of Alberta, Edmonton, Alberta, T6G 2E1, Canada}
\affiliation{Erlangen Centre for Astroparticle Physics, Friedrich-Alexander-Universit{\"a}t Erlangen-N{\"u}rnberg, D-91058 Erlangen, Germany}
\affiliation{Physik-department, Technische Universit{\"a}t M{\"u}nchen, D-85748 Garching, Germany}
\affiliation{D{\'e}partement de physique nucl{\'e}aire et corpusculaire, Universit{\'e} de Gen{\`e}ve, CH-1211 Gen{\`e}ve, Switzerland}
\affiliation{Dept. of Physics and Astronomy, University of Gent, B-9000 Gent, Belgium}
\affiliation{Dept. of Physics and Astronomy, University of California, Irvine, CA 92697, USA}
\affiliation{Karlsruhe Institute of Technology, Institute for Astroparticle Physics, D-76021 Karlsruhe, Germany}
\affiliation{Karlsruhe Institute of Technology, Institute of Experimental Particle Physics, D-76021 Karlsruhe, Germany}
\affiliation{Dept. of Physics, Engineering Physics, and Astronomy, Queen's University, Kingston, ON K7L 3N6, Canada}
\affiliation{Department of Physics {\&} Astronomy, University of Nevada, Las Vegas, NV 89154, USA}
\affiliation{Nevada Center for Astrophysics, University of Nevada, Las Vegas, NV 89154, USA}
\affiliation{Dept. of Physics and Astronomy, University of Kansas, Lawrence, KS 66045, USA}
\affiliation{Centre for Cosmology, Particle Physics and Phenomenology - CP3, Universit{\'e} catholique de Louvain, Louvain-la-Neuve, Belgium}
\affiliation{Department of Physics, Mercer University, Macon, GA 31207-0001, USA}
\affiliation{Dept. of Astronomy, University of Wisconsin{\textemdash}Madison, Madison, WI 53706, USA}
\affiliation{Dept. of Physics and Wisconsin IceCube Particle Astrophysics Center, University of Wisconsin{\textemdash}Madison, Madison, WI 53706, USA}
\affiliation{Institute of Physics, University of Mainz, Staudinger Weg 7, D-55099 Mainz, Germany}
\affiliation{Department of Physics, Marquette University, Milwaukee, WI 53201, USA}
\affiliation{Institut f{\"u}r Kernphysik, Universit{\"a}t M{\"u}nster, D-48149 M{\"u}nster, Germany}
\affiliation{Bartol Research Institute and Dept. of Physics and Astronomy, University of Delaware, Newark, DE 19716, USA}
\affiliation{Dept. of Physics, Yale University, New Haven, CT 06520, USA}
\affiliation{Columbia Astrophysics and Nevis Laboratories, Columbia University, New York, NY 10027, USA}
\affiliation{Dept. of Physics, University of Oxford, Parks Road, Oxford OX1 3PU, United Kingdom}
\affiliation{Dipartimento di Fisica e Astronomia Galileo Galilei, Universit{\`a} Degli Studi di Padova, I-35122 Padova PD, Italy}
\affiliation{Dept. of Physics, Drexel University, 3141 Chestnut Street, Philadelphia, PA 19104, USA}
\affiliation{Physics Department, South Dakota School of Mines and Technology, Rapid City, SD 57701, USA}
\affiliation{Dept. of Physics, University of Wisconsin, River Falls, WI 54022, USA}
\affiliation{Dept. of Physics and Astronomy, University of Rochester, Rochester, NY 14627, USA}
\affiliation{Department of Physics and Astronomy, University of Utah, Salt Lake City, UT 84112, USA}
\affiliation{Dept. of Physics, Chung-Ang University, Seoul 06974, Republic of Korea}
\affiliation{Oskar Klein Centre and Dept. of Physics, Stockholm University, SE-10691 Stockholm, Sweden}
\affiliation{Dept. of Physics and Astronomy, Stony Brook University, Stony Brook, NY 11794-3800, USA}
\affiliation{Dept. of Physics, Sungkyunkwan University, Suwon 16419, Republic of Korea}
\affiliation{Institute of Physics, Academia Sinica, Taipei, 11529, Taiwan}
\affiliation{Dept. of Physics and Astronomy, University of Alabama, Tuscaloosa, AL 35487, USA}
\affiliation{Dept. of Astronomy and Astrophysics, Pennsylvania State University, University Park, PA 16802, USA}
\affiliation{Dept. of Physics, Pennsylvania State University, University Park, PA 16802, USA}
\affiliation{Dept. of Physics and Astronomy, Uppsala University, Box 516, SE-75120 Uppsala, Sweden}
\affiliation{Dept. of Physics, University of Wuppertal, D-42119 Wuppertal, Germany}
\affiliation{Deutsches Elektronen-Synchrotron DESY, Platanenallee 6, D-15738 Zeuthen, Germany}

\author{R. Abbasi}
\affiliation{Department of Physics, Loyola University Chicago, Chicago, IL 60660, USA}
\author{M. Ackermann}
\affiliation{Deutsches Elektronen-Synchrotron DESY, Platanenallee 6, D-15738 Zeuthen, Germany}
\author{J. Adams}
\affiliation{Dept. of Physics and Astronomy, University of Canterbury, Private Bag 4800, Christchurch, New Zealand}
\author{S. K. Agarwalla}
\thanks{also at Institute of Physics, Sachivalaya Marg, Sainik School Post, Bhubaneswar 751005, India}
\affiliation{Dept. of Physics and Wisconsin IceCube Particle Astrophysics Center, University of Wisconsin{\textemdash}Madison, Madison, WI 53706, USA}
\author{J. A. Aguilar}
\affiliation{Universit{\'e} Libre de Bruxelles, Science Faculty CP230, B-1050 Brussels, Belgium}
\author{M. Ahlers}
\affiliation{Niels Bohr Institute, University of Copenhagen, DK-2100 Copenhagen, Denmark}
\author{J.M. Alameddine}
\affiliation{Dept. of Physics, TU Dortmund University, D-44221 Dortmund, Germany}
\author{S. Ali}
\affiliation{Dept. of Physics and Astronomy, University of Kansas, Lawrence, KS 66045, USA}
\author{N. M. Amin}
\affiliation{Bartol Research Institute and Dept. of Physics and Astronomy, University of Delaware, Newark, DE 19716, USA}
\author{K. Andeen}
\affiliation{Department of Physics, Marquette University, Milwaukee, WI 53201, USA}
% ...

\date{\today}

\collaboration{IceCube Collaboration}
\noaffiliation

\begin{abstract}
Dark matter is approximately five times more abundant than baryonic matter in the universe, but its physical nature continues to elude physicists. One potential candidate for dark matter is a weakly-interacting massive particle (WIMP), which is predicted by various extensions to the Standard Model (SM) of particle physics. After becoming gravitationally bound in cosmic structures, WIMPs can self-annihilate and produce SM particles including neutrinos, which are observable by detectors like IceCube. We present a search for neutrinos from low-mass $(\leq 300 \, \mathrm{GeV})$ WIMP annihilation in dwarf spheroidal galaxies with over seven years of IceCube livetime. We find no statistically significant evidence of neutrinos produced by WIMP annihilation, and therefore set upper limits on the velocity-averaged annihilation cross section $\left<\sigma v\right>$. Our strongest upper limits at the 90\% confidence level are $\mathcal{O}\!\left(10^{-22} \, \mathrm{{cm}^{3} \, s^{-1}}\right)$ for WIMP annihilation directly into neutrino-antineutrino pairs. For our least sensitive channel, the corresponding limits are $\mathcal{O}\!\left(10^{-20} \, \mathrm{{cm}^{3} \, s^{-1}}\right)$, which is an improvement of over two orders of magnitude over previous IceCube limits from dwarf galaxies at the upper end of our mass range.
\end{abstract}

\maketitle

\section{Introduction} \label{sec:introduction}

Several astronomical observations provide indirect evidence for dark matter (DM) via its gravitational interactions, including galaxy rotation curves, galaxy and galaxy cluster mergers, and cosmological measurements~\cite{Bertone2018,Leane2018,Cirelli2024}. DM is believed to make up approximately five times as much of the universe's mass-energy density as baryonic matter~\cite{Ade2014}, though physicists continue to probe its fundamental nature. Its lack of electromagnetic interactions makes detection difficult, requiring either very sensitive measurements of rare DM interactions in shielded underground detectors such as LUX-ZEPLIN (LZ)~\cite{Aalbers2023,Aalbers2024a,Aalbers2024b}, PandaX~\cite{Tan2016,Fu2017,Fu2018,Cui2017}, and XENON~\cite{Angle2008,Aprile2010,Aprile2011,Aprile2017}; direct DM production in high-energy colliders such as the LHC~\cite{Kahlhoefer2017}; or indirect searches for the byproducts of DM annihilation or decay.

One leading candidate particle for DM is a stable weakly-interacting massive particle (WIMP), which is predicted by several extensions to the SM~\cite{Jungman1996,Bertone2005}. WIMPs are believed to interact via the gravitational force, and potentially via forces even weaker than the weak scale~\cite{Kamionkowski1997}. WIMPs may annihilate and produce SM particle-antiparticle pairs in a process given by $\chi\chi \to b\bar{b}, W^{+}W^{-}, \mu^{+}\mu^{-}, \tau^{+}\tau^{-}, \nu_{e,\mu,\tau} \bar{\nu}_{e,\mu,\tau}, \gamma\gamma, \dots$, where $\chi$ represents a DM particle. Notably, this process would produce stable SM particles like neutrinos and gamma-rays, either directly or indirectly via decays and secondary interactions, which would propagate to Earth. This means that observatories can search for excesses over astrophysical backgrounds of neutrinos and gamma-rays produced in WIMP annihilation. Such searches have been performed by a variety of gamma-ray observatories such as Fermi-LAT~\cite{Ackermann2014,Ackermann2015,Albert2017}, HAWC~\cite{Albert2018}, VERITAS~\cite{Acciari2010,Aliu2012,Aliu2015,Archambault2017,Acharyya2023,Acharyya2024}, MAGIC~\cite{MAGIC2016,Ahnen2018,Acciari2022}, H.E.S.S.~\cite{Abramowski2014,Cirelli2018,Abdallah2020}, and LHAASO~\cite{Cao2024}. Neutrino observatories like IceCube~\cite{Aartsen2017a,Aartsen2024}, ANTARES~\cite{Ageron2011}, KM3NeT~\cite{Adrian-Martinez2016}, and Baikal~\cite{Avrorin2015} have also performed such searches, targeting DM in dwarf galaxies and nearby galaxy clusters~\cite{Aartsen2013}, the Sun~\cite{Aartsen2017b,Aartsen2019b,Abbasi2022b,Aiello2025}, the Galactic halo~\cite{Aartsen2016}, and the Galactic center~\cite{Avrorin2016,Aartsen2017c,Albert2020a,Albert2020b,Iovine2021,Abbasi2023,Aiello2025}.

The expected flux of neutrinos from WIMP annihilation is,

\begin{equation} \label{eq:wimp_annihilation}
\begin{split}
\phi\!\left(\Delta\Omega\right) & = \frac{1}{8\pi} \frac{\left<\sigma v\right>}{m_{\chi}^{2}} \int_{E_{\text{min}}}^{E_{\text{max}}} {\frac{dN}{dE_{\nu}} \ dE_{\nu}} \\
 & \times \int_{\text{l.o.s}} {\int_{\Delta\Omega} {\rho_{\chi}^{2}\!\left(r\right) \ d\Omega'} \ d\ell'},
\end{split}
\end{equation}

\noindent where $\phi$ represents the flux of neutrinos at the detector, $\Delta\Omega$ represents the solid angle of sky being observed, $\left<\sigma v\right>$ represents the velocity-averaged WIMP annihilation cross section, $m_{\chi}$ represents the WIMP mass, $\frac{dN}{dE_{\nu}}$ represents the spectrum of neutrinos produced from WIMP annihilation (given a particular WIMP mass and annihilation channel), and $\rho_{\chi}(r)$ represents the DM density profile of the source being observed. The double integral over the line of sight and the solid angle in Eq.~(\ref{eq:wimp_annihilation}) is typically referred to as the J-factor $J(\theta)$, where $\theta$ is the opening half-angle of the source such that $\Delta\Omega = \pi (1 - \cos(\theta))$, which approaches $\pi \theta^{2}$ in the limit of small $\theta$.

In this analysis, we focus on a set of dwarf spheroidal galaxies (dSphs) as potential sites of DM annihilation. This is because dSphs are the most DM-dominated structures known, and because they have very little high-energy astrophysical activity, leading to a negligible SM astrophysical neutrino background~\cite{Evans2004,Sandick2010}. For our analysis, we assume that the DM density profile $\rho_{\chi}(r)$ follows the Navarro-Frenk-White (NFW) profile:

\begin{equation} \label{eq:NFW_profile}
\rho_{\text{NFW}}(r) \propto \frac{1}{r \left(1 + r/r_{s}\right)^{2}},
\end{equation}

\noindent where $r_{s}$ is the scale radius as defined in~\cite{Navarro1996}. We recognize that the distribution of DM within dSphs remains under study, with some suggesting that the DM radial profiles may be more cored than cusped; see, e.g.,~\cite{Salucci2019,Cirelli2024} for a review. However, we note that this work is largely insensitive to changes in the DM density profile at small radial distances due to the large angular resolution of the neutrino events in our dataset, as will be discussed in Sec.~\ref{sec:icecube}.

The remainder of this paper is outlined as follows. In Sec.~\ref{sec:icecube}, we give an overview of the IceCube detector and the neutrino data used in this analysis. In Sec.~\ref{sec:external_data}, we describe the additional inputs in this analysis; namely, the WIMP annihilation spectra (Sec.~\ref{sec:wimp_annihilation_spectra}) and the dSphs (Sec.~\ref{sec:dwarf_spheroidals}). In Sec.~\ref{sec:analysis_methods}, we describe our analysis methods, beginning with the likelihood function (Sec.~\ref{sec:likelihood}) and the construction of probability distribution functions (PDFs, Sec.~\ref{sec:background_signal_pdfs}). In Sec.~\ref{sec:results}, we discuss our results. In Sec.~\ref{sec:systematic_uncertainties}, we discuss systematic uncertainties in our analysis. In Sec.~\ref{sec:conclusions}, we conclude and discuss implications for future work. Appendix~\ref{apx:pdfs} contains sample sets of PDFs, and Appendix~\ref{apx:post_unblinding_checks} contains additional cross-checks.

%\FloatBarrier

\section{IceCube} \label{sec:icecube}

IceCube is an in-ice Cherenkov detector comprising $1 \, \mathrm{{km}^{3}}$ of Antarctic ice below the South Pole. The detector consists of 5,160 digital optical modules (DOMs) arranged on 86 cables ("strings"), deployed at a depth of $1.5 - 2.5 \, \mathrm{km}$~\cite{Abbasi2009,Abbasi2010,Halzen2010,Aartsen2014,Aartsen2017a,Aartsen2024}. Each DOM has a downward-facing photomultiplier tube encased in a glass pressure housing. Seventy-eight of the strings contain 60 DOMs with a $17 \, \mathrm{m}$ vertical spacing, and an approximate average string spacing of $125 \, \mathrm{m}$. The remaining 8 strings contain 60 high-quantum efficiency DOMs with a $7 \, \mathrm{m}$ vertical spacing, and an approximate average string spacing of $75 \, \mathrm{m}$. The strings are arranged on a hexagonal grid, with the high-density strings placed at the center of the detector. This high-instrument density region is referred to as DeepCore, which is designed to study low-energy $\left(\mathcal{O}\!\left(10 \, \mathrm{GeV}\right)\right)$ neutrino events~\cite{Abbasi2012}.

We make use of a low-energy $\left(0.5 \lesssim \log\!\left(E_{\nu}/\mathrm{GeV}\right) \lesssim 2.5\right)$ neutrino dataset consisting of $\sim\!1.67$ million events collected between 2011 and 2017, comprising $\sim\!2600$ days of detector livetime. This dataset was initially produced to study $\nu_{\mu} \to \nu_{\tau}$ oscillations of low-energy atmospheric neutrinos~\cite{Aartsen2019a}, but has since been reprocessed for DM searches~\cite{Abbasi2022b}. The median angular resolution of the dataset is energy-dependent, with a resolution of $\approx 48^{\circ}$ at $5 \, \mathrm{GeV}$ and $\approx 10^{\circ}$ at $100 \, \mathrm{GeV}$.

%\FloatBarrier

\section{Dark Matter in Search Targets} \label{sec:external_data}

\subsection{WIMP Annihilation Spectra} \label{sec:wimp_annihilation_spectra}

The spectra of neutrinos produced in DM annihilation [Eq.~(\ref{eq:wimp_annihilation})] are calculated using the Poor Particle Physicist Cookbook (PPPC)~\cite{Cirelli2011,Cirelli2012}. PPPC simulates the production of particles from DM annihilation given a DM particle mass and an annihilation channel. In accordance with~\cite{Aartsen2013}, we investigate five different annihilation channels: bottom quarks $b\bar{b}$, W bosons $W^{+}W^{-}$, muons $\mu^{+}\mu^{-}$, taus $\tau^{+}\tau^{-}$, and neutrinos $\nu\bar{\nu}$. Note that this is an average neutrino channel; we do not distinguish between neutrino flavors in this analysis, mainly due to the difficulty of accurately reconstructing flavor in this energy range. These annihilation spectra include electroweak corrections, which are primarily important for the $W^{+}W^{-}$ and $\nu\bar{\nu}$ channels~\cite{Ciafaloni2011}.

\begin{figure*}[htb!]
\centering
\begin{subfigure}{0.45\textwidth}
\includegraphics[width=\linewidth,height=5cm]{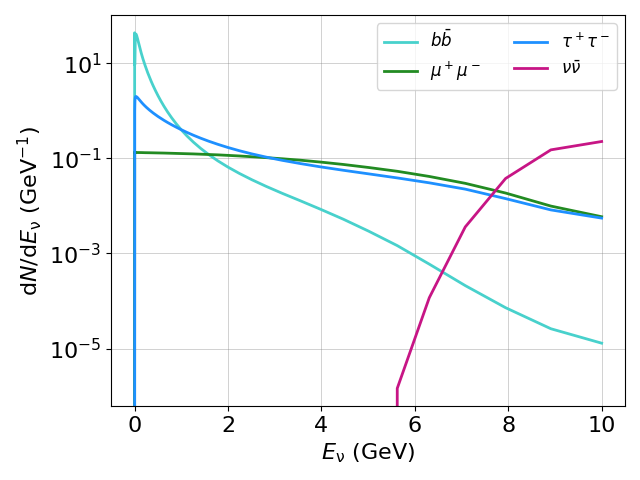}
\end{subfigure}
\begin{subfigure}{0.45\textwidth}
\includegraphics[width=\linewidth,height=5cm]{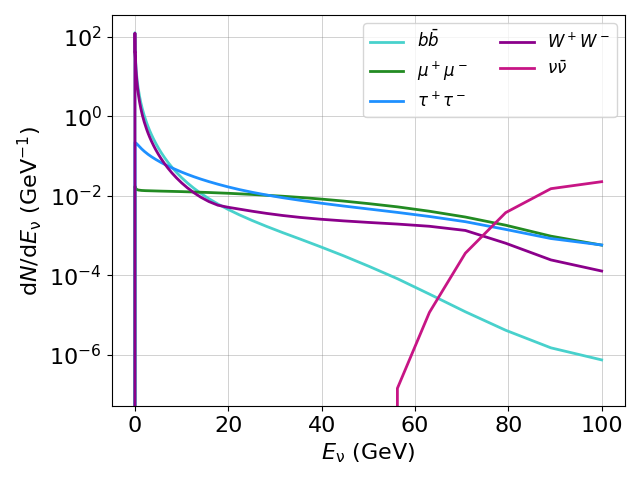}
\end{subfigure}
\begin{subfigure}{0.45\textwidth}
\includegraphics[width=\linewidth,height=5cm]{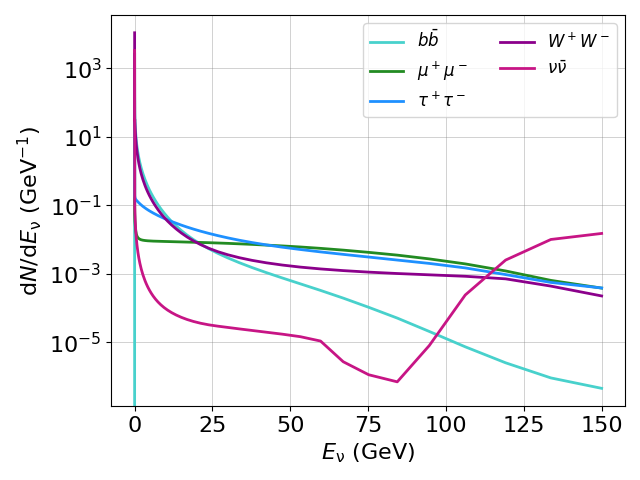}
\end{subfigure}
\begin{subfigure}{0.45\textwidth}
\includegraphics[width=\linewidth,height=5cm]{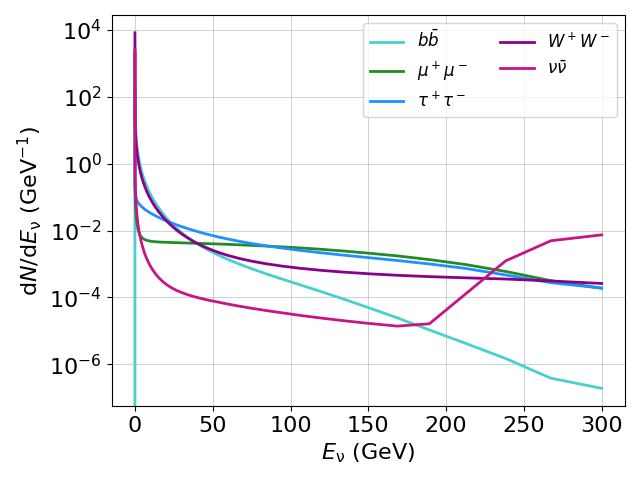}
\end{subfigure}
\caption{\label{fig:annihilation_spectra}WIMP annihilation spectra for WIMP masses of $10 \, \mathrm{GeV}$ (upper left), $100 \, \mathrm{GeV}$ (upper right), $150 \, \mathrm{GeV}$ (lower left), and $300 \, \mathrm{GeV}$ (lower right). Note the physicality constraint for the $W^{+}W^{-}$ channel ($m_{\chi} \geq 90 \, \mathrm{GeV}$), and the production of the low-energy tail in the $\nu\bar{\nu}$ spectra in the bottom row.}
\end{figure*}

We apply neutrino oscillations to the spectra to generate the spectra as observed at Earth~\cite{Blennow2008}. Oscillations are parametrized using the Pontecorvo-Maki-Nakagawa-Sakata matrix~\cite{Pontecorvo1958,Maki1962,Chau1984}:

\begin{widetext}
\begin{equation} \label{eq:PMNS_matrix}
U = \begin{pmatrix} c_{12} c_{23} & s_{12} c_{13} & s_{13} e^{i\delta_{\text{CP}}} \\ -s_{12} c_{23} - c_{12} s_{23} s_{13} e^{i\delta_{\text{CP}}} & c_{12} c_{23} - s_{12} s_{23} s_{13} e^{i\delta_{\text{CP}}} & s_{23} c_{13} \\ s_{12} s_{23} - c_{12} c_{23} s_{13} e^{i\delta_{\text{CP}}} & -c_{12} s_{23} - s_{12} c_{23} s_{13} e^{i\delta_{\text{CP}}} & c_{23} c_{13} \end{pmatrix},
\end{equation}
\end{widetext}

\noindent where $c_{ij}/s_{ij} = \cos(\theta_{ij})/\sin(\theta_{ij})$, $\theta_{ij}$ represent the neutrino mixing angles, and $\delta_{CP}$ represents the CP-violating phase. We assume the oscillation parameters $\theta_{12} = 0.57596$, $\theta_{23} = 0.8552$, $\theta_{13} = 0.1296$~\cite{Beringer2012} and choose $\delta_{\text{CP}} = 0$ since we do not distinguish between neutrinos and antineutrinos in this analysis. We investigated the impact of using a more recent set of mixing angles from~\cite{Abe2021,Esteban2024} and found the annihilation spectra to be consistent within 0.5\%. Discrepancies at this scale will have minimal impact on our results and are well below the uncertainties on the J-factors (see Sec.~\ref{sec:dwarf_spheroidals}).

Since we are in the oscillation-averaged regime due to propagation over a large distance, the probability of a neutrino oscillating between flavor eigenstates $a$ and $b$ is given by

\begin{equation}
\mathcal{P}\!\left(a \to b\right) = \sum_{k=1}^{3} {\text{Re}\left[U_{ak}^{*} U_{bk}^{ } U_{ak}^{ } U_{bk}^{*}\right]},
\end{equation}

\noindent where $k$ runs over all mass eigenstates and $U^{*}$ is the complex conjugate of $U$.

A sample of our annihilation spectra are shown in Fig.~\ref{fig:annihilation_spectra}. The $W^{+}W^{-}$ channel is not shown in the upper left panel due to physicality constraints $\left(\chi\chi \to W^{+}W^{-}\right.$ only for $\left.m_{\chi} \ge 90 \, \mathrm{GeV}\right)$. Most channels are well-approximated by power laws, with $b\bar{b}$ being a soft spectrum and $\mu^{+}\mu^{-}$ and $\tau^{+}\tau^{-}$ being harder spectra. There are two effects that contribute to the distinct shapes of the $\nu\bar{\nu}$ spectra:

\begin{itemize}
    \item[\textbullet] electroweak corrections introduce quantum effects that lead to the broadening of the peak down to $\sim\!60\%$ of the WIMP mass; and
    \item[\textbullet] at higher WIMP masses, $W$ and $Z$ bosons can be produced "on-shell" during the annihilation, which then interact and decay to produce the tail of low-energy neutrinos. This effect appears for $m_{\chi} \gtrsim 125 \, \mathrm{GeV}$.
\end{itemize}

\begin{figure*}[htb!]
\centering
\includegraphics[width=0.83\linewidth]{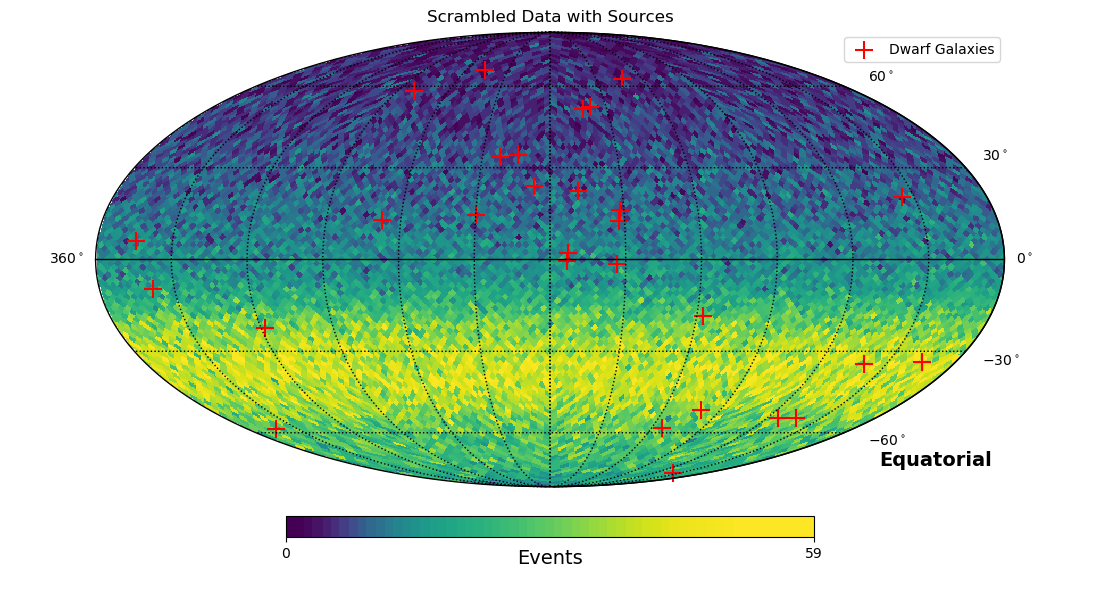}
\caption{\label{fig:skymap} The locations of dSphs (red crosses) plotted over a skymap of one year of neutrino data with randomized right ascensions of events. The colorbar is events per pixel and the pixel sizes correspond to $N_{\text{side}} = 32$ \cite{Gorski2005}.}
\end{figure*}

%\FloatBarrier

\subsection{Dwarf Spheroidals} \label{sec:dwarf_spheroidals}

We adopt a set of 29 dSphs used as sources of WIMP annihilation in this analysis, with 25 from~\cite{Boddy2020} and 4 additional sources from~\cite{Baring2016,Evans2016}. Their coordinates and J-factors are provided in Table~\ref{tab:dwarfs} and their locations are plotted in Fig.~\ref{fig:skymap}. Unless otherwise noted, we assume the J-factor with the maximum opening half-angle available for each source~\cite{Boddy2020,Baring2016,Evans2016}, representing a "best-case" scenario in which our expected flux is maximized according to Eq.~(\ref{eq:wimp_annihilation}); this maximum opening half-angle is either $10^{\circ}$ or $0.5^{\circ}$ when the former is not available. The impact of different J-factor assumptions is discussed under systematic uncertainties (Sec.~\ref{sec:systematic_uncertainties}).

\begin{table*}[htb!]
{\renewcommand{\arraystretch}{1.2}%
\begin{ruledtabular}
\begin{tabular}{cddddc}
    Source & \multicolumn{1}{c}{\text{RA}} & \multicolumn{1}{c}{\text{Dec}} & \multicolumn{1}{c}{$\log_{10}\left(J(0.5\degree)\right)$} & \multicolumn{1}{c}{$\log_{10}\left(J(10\degree)\right)$} & Coordinate \\
     & \multicolumn{1}{c}{$\left[\degree\right]$} & \multicolumn{1}{c}{$\left[\degree\right]$} & \multicolumn{1}{c}{$\left[\mathrm{GeV}^{2} \, \mathrm{cm}^{-5}\right]$} & \multicolumn{1}{c}{$\left[\mathrm{GeV}^{2} \, \mathrm{cm}^{-5}\right]$} & Reference \\
    \colrule
    Aquarius II       & 338.48 &  -9.33 & 18.39^{+0.65}_{-0.58} & 18.48^{+0.72}_{-0.62} & \cite{Torrealba2016b} \\
    Boötes I          & 210.05 &  14.49 & 18.19^{+0.30}_{-0.28} & 18.39^{+0.45}_{-0.38} & \cite{Ackermann2014} \\
    Canes Venatici I  & 202.02 &  33.56 & 17.43^{+0.16}_{-0.15} & 17.49^{+0.24}_{-0.17} & \cite{Simon2007} \\
    Canes Venatici II & 194.29 &  34.32 & 17.82^{+0.48}_{-0.47} & 17.92^{+0.55}_{-0.53} & \cite{Simon2007} \\
    Carina I          & 100.43 & -50.95 & 17.83^{+0.09}_{-0.09} & 17.88^{+0.11}_{-0.10} & \cite{Ackermann2014} \\
    Carina II         & 114.11 & -58.00 & 18.37^{+0.54}_{-0.52} & 18.57^{+0.66}_{-0.59} & \cite{Torrealba2018} \\
    Coma Berenices    & 186.75 &  23.90 & 19.01^{+0.36}_{-0.36} & 19.25^{+0.57}_{-0.50} & \cite{Simon2007} \\
    Crater II         & 117.31 & -18.41 & 15.35^{+0.27}_{-0.25} & 15.56^{+0.25}_{-0.23} & \cite{Torrealba2016b} \\
    Draco             & 260.09 &  57.94 & 18.82^{+0.12}_{-0.12} & 18.96^{+0.20}_{-0.16} & \cite{Ackermann2014} \\
    Fornax            &  39.94 & -34.44 & 18.09^{+0.10}_{-0.10} & 18.11^{+0.10}_{-0.09} & \cite{Ackermann2014} \\
    Hercules          & 247.76 &  12.79 & 17.29^{+0.51}_{-0.52} & 17.35^{+0.58}_{-0.54} & \cite{Simon2007} \\
    Horologium I      &  43.87 & -54.12 & 19.17^{+0.80}_{-0.70} & 19.28^{+0.87}_{-0.80} & \cite{Jerjen2018} \\
    Hydrus            &  37.39 & -79.31 & 18.65^{+0.32}_{-0.31} & 18.93^{+0.57}_{-0.47} & \cite{Koposov2018} \\
    Leo I             & 152.11 &  12.31 & 17.61^{+0.13}_{-0.11} & 17.64^{+0.19}_{-0.12} & \cite{Mateo2008} \\
    Leo II            & 168.34 &  22.13 & 17.66^{+0.16}_{-0.15} & 17.66^{+0.16}_{-0.15} & \cite{Ackermann2014} \\
    Reticulum II      &  53.93 & -54.06 & 18.94^{+0.38}_{-0.38} & 19.16^{+0.64}_{-0.53} & \cite{Koposov2015} \\
    Sagittarius II    & 298.17 & -22.07 & 17.35^{+1.36}_{-0.91} & 17.48^{+1.23}_{-0.79} & \cite{Laevens2015} \\
    Sculptor          &  15.03 & -33.67 & 18.58^{+0.05}_{-0.05} & 18.63^{+0.05}_{-0.05} & \cite{Ackermann2014} \\
    Segue 1           & 151.76 &  16.07 & 19.00^{+0.48}_{-0.68} & 19.12^{+0.63}_{-0.68} & \cite{Simon2011} \\
    Sextans           & 153.28 &  -1.60 & 17.75^{+0.12}_{-0.11} & 17.87^{+0.12}_{-0.10} & \cite{Ackermann2014} \\
    Tucana II         & 343.06 & -58.57 & 18.93^{+0.56}_{-0.50} & 19.13^{+0.65}_{-0.56} & \cite{Bechtol2015} \\
    Ursa Major I      & 158.72 &  51.92 & 18.33^{+0.28}_{-0.28} & 18.40^{+0.37}_{-0.32} & \cite{Simon2007} \\
    Ursa Major II     & 132.88 &  63.13 & 19.44^{+0.41}_{-0.39} & 19.72^{+0.54}_{-0.49} & \cite{Simon2007} \\
    Ursa Minor        & 227.24 &  67.24 & 18.76^{+0.12}_{-0.11} & 18.80^{+0.11}_{-0.11} & \cite{Ackermann2014} \\
    Willman 1         & 162.35 &  51.03 & 19.36^{+0.52}_{-0.46} & 19.46^{+0.73}_{-0.52} & \cite{Ackermann2014} \\
    \colrule
    Leo IV            & 173.24 &  -0.53 & 16.12^{+0.71}_{-1.14} & \multicolumn{1}{c}{-} & \cite{Simon2007} \\
    Leo V             & 172.79 &   2.18 & 15.86^{+0.46}_{-0.47} & \multicolumn{1}{c}{-} & \cite{Ackermann2014} \\
    Segue 2           &  34.81 &  20.22 & 15.89^{+0.56}_{-0.37} & \multicolumn{1}{c}{-} & \cite{Ackermann2014} \\
    \colrule
    Pisces II         & 344.62 &   5.95 & 17.90^{+1.14}_{-0.90} & \multicolumn{1}{c}{-} & \cite{Ackermann2014}
\end{tabular}
\end{ruledtabular}
\caption{\label{tab:dwarfs}Data of dSphs used in analysis. Right ascension (RA) and declination are given in equatorial coordinates. All J-factors are taken from~\cite{Boddy2020}, except those for Leo IV, Leo V, and Segue 2 taken from~\cite{Baring2016} and Pisces II from~\cite{Evans2016}. Coordinate references are typically taken from~\cite{Ackermann2014} or references therein.}
}
\end{table*}

%\FloatBarrier

\section{Analysis Methods} \label{sec:analysis_methods}

\subsection{Likelihood} \label{sec:likelihood}

In this analysis, we fit for the most-likely number of signal events in our data using the unbinned likelihood ratio method~\cite{Braun2008,Abbasi2022a}. For multiple ("stacked") sources, this log likelihood (LLH) is given by

\begin{widetext}
\begin{equation} \label{eq:LLH_stacked}
\ln\!\left(\mathcal{L}\left(n_{s}\right)\right) = \sum_{k=1}^{N_{\text{sources}}} {\sum_{i=1}^{N_{\text{events}}} {\left[\frac{n_{s}}{N_{\text{events}}} \mathcal{S}_{k}\!\left(\psi_{i}\right) w_{k} + \left(1 - \frac{n_{s}}{N_{\text{events}}}\right) \mathcal{B}_{k}\!\left(\psi_{i}\right)\right]}},
\end{equation}
\end{widetext}

\noindent where $n_{s}$ is the number of signal events, $w_{k}$ represents the weight for each source, and $\mathcal{S}_{k}\!\left(\psi_{i}\right)/\mathcal{B}_{k}\!\left(\psi_{i}\right)$ represent the signal\slash background probability distribution functions (PDFs) for source $k$ evaluated for event $i$, respectively, where $\psi_{i}$ represents the angular separation between the event and the source. For our analysis, we choose

\begin{equation} \label{eq:source_weight}
w_{k} = \frac{J_{k}}{\sum_{\ell} {J_{\ell}}},
\end{equation}

\noindent since this represents each source's expected contribution to our signal. The best-fit value of $n_{s}$ is found by maximizing the LLH. We also calculate a summary statistic called the test statistic, defined as

\begin{equation} \label{eq:TS}
\mathrm{TS} = -2 \left[\ln\!\left(\mathcal{L}\left(n_{s}\right)\right) - \ln\!\left(\mathcal{L}\left(n_{s} = 0\right)\right)\right],
\end{equation}

\noindent such that the test statistic describes our fit relative to a background-only fit.

%\FloatBarrier

\subsection{Background and Signal PDFs} \label{sec:background_signal_pdfs}

For $\mathcal{B}_{k}\!\left(\psi_{i}\right)$ and $\mathcal{S}_{k}\!\left(\psi_{i}\right)$, we must construct PDFs, which are functions of the angular separation from the source for a given range of reconstructed neutrino energies. These describe the probability of a neutrino event having a given angular separation from source $k$.

First, we calculate the optimal energy ranges used to construct the PDFs. This energy range is selected to optimize the ratio $\frac{S}{\sqrt{B}}$, where $S$ represents the signal weights and $B$ represents the background weights. These energy ranges are listed in Table~\ref{tab:energy_ranges}. These energy ranges are used as a cut on our sample for generating the PDFs, effectively collapsing the energy dimension such that our PDFs are one-dimensional and only depend on the angular separation.

Background PDFs are generated for each channel-mass-source combination, as a given neutrino will have different probabilities of being considered a background event for different sources. For each neutrino that passes the energy range cut for a given channel-mass combination, we resample this neutrino $N$ times with random RA values. The choice of $N$ is somewhat arbitrary and just needs to be large enough to generate smooth PDFs; we adopt $N = 30$ in accordance with~\cite{Abbasi2022b}. This resampling approach is valid because we expect the background to be isotropic in RA due to the Earth's rotation and the high uptime of the detector ($\sim\!98\%$). We then calculate the angular separations between the resampled neutrinos and each source. The normalized histogram of all the angular separations comprises a background PDF. This resampling procedure results in smooth PDFs as a function of both WIMP mass and angular separation from the source, as shown in the sample background PDFs in Figs.~\ref{fig:background_pdfs_LeoIV} and \ref{fig:background_pdfs_UrsaMajorI}. We show background PDFs for two different sources to highlight the effect of the source declination on the shapes of the PDFs.

\begin{table*}[ht!]
\begin{ruledtabular}
\begin{tabular}{cccccc}
    WIMP Mass & $b\bar{b}$ & $W^{+}W^{-}$ & $\mu^{+}\mu^{-}$ & $\tau^{+}\tau^{-}$ & $\nu\bar{\nu}$ \\
    $\left[\mathrm{GeV}\right]$ & $\left(\left[\mathrm{GeV}\right], \left[\mathrm{GeV}\right]\right)$ & $\left(\left[\mathrm{GeV}\right], \left[\mathrm{GeV}\right]\right)$ & $\left(\left[\mathrm{GeV}\right], \left[\mathrm{GeV}\right]\right)$ & $\left(\left[\mathrm{GeV}\right], \left[\mathrm{GeV}\right]\right)$ & $\left(\left[\mathrm{GeV}\right], \left[\mathrm{GeV}\right]\right)$ \\
    \colrule
      5 & -        & -          & (0, 9)     & (0, 9)    & (1, 12) \\
     10 & (0, 12)  & -          & (0, 16)    & (0, 16)   & (0, 21) \\
     20 & (0, 18)  & -          & (3, 30)    & (0, 30)   & (12, 39) \\
     30 & (0, 25)  & -          & (6, 43)    & (4, 44)   & (19, 53) \\
     50 & (0, 38)  & -          & (16, 49)   & (12, 71)  & (34, 83) \\
     80 & (0, 57)  & -          & (24, 107)  & (20, 108) & (56, 128) \\
     90 & -        & (27, 108)  & -          & -         & - \\
    100 & (0, 73)  & (31, 128)  & (31, 129)  & (27, 138) & (70, 159) \\
    150 & (0, 108) & (54, 207)  & (48, 199)  & (40, 205) & (113, 234) \\
    200 & (3, 152) & (82, 298)  & (66, 266)  & (57, 272) & (148, 298) \\
    300 & (6, 256) & (120, 448) & (100, 389) & (92, 389) & (234, 441)
\end{tabular}
\end{ruledtabular}
\caption{\label{tab:energy_ranges}Energy ranges used to construct PDFs. Note the physicality constraints on the $b\bar{b}$ $(m_{\chi} \ge 10 \ \mathrm{GeV})$ and $W^{+}W^{-}$ $(m_{\chi} \ge 90 \ \mathrm{GeV})$ channels. The mass point $m_{\chi} = 90 \ \mathrm{GeV}$ replaces $80 \ \mathrm{GeV}$ for the $W^{+}W^{-}$ channel due to this constraint.}
\end{table*}

\begin{table*}[htb!]
{\renewcommand{\arraystretch}{1.15}%
\begin{ruledtabular}
\begin{tabular}{cccccc}
      WIMP Mass & $b\bar{b}$ & $W^{+}W^{-}$ & $\mu^{+}\mu^{-}$ & $\tau^{+}\tau^{-}$ & $\nu\bar{\nu}$ \\
      $\left[\mathrm{GeV}\right]$ & $\left[\mathrm{cm}^{3} \, \mathrm{s}^{-1}\right]$ & $\left[\mathrm{cm}^{3} \, \mathrm{s}^{-1}\right]$ & $\left[\mathrm{cm}^{3} \, \mathrm{s}^{-1}\right]$ & $\left[\mathrm{cm}^{3} \, \mathrm{s}^{-1}\right]$ & $\left[\mathrm{cm}^{3} \, \mathrm{s}^{-1}\right]$ \\
      \colrule
        5 & - & - & $1.456 \times 10^{-20}$ & $2.015 \times 10^{-20}$ & $3.518 \times 10^{-21}$ \\
       10 & $3.440 \times 10^{-19}$ & - & $4.732 \times 10^{-21}$ & $6.692 \times 10^{-21}$ & $5.472 \times 10^{-22}$ \\
       20 & $7.257 \times 10^{-20}$ & - & $2.718 \times 10^{-21}$ & $4.305 \times 10^{-21}$ & $2.016 \times 10^{-22}$ \\
       30 & $5.287 \times 10^{-20}$ & - & $1.642 \times 10^{-21}$ & $2.504 \times 10^{-21}$ & $3.075 \times 10^{-22}$ \\
       50 & $2.797 \times 10^{-20}$ & - & $1.252 \times 10^{-21}$ & $1.157 \times 10^{-21}$ & $2.312 \times 10^{-22}$ \\
       80 & $2.163 \times 10^{-20}$ & - & $1.387 \times 10^{-21}$ & $1.374 \times 10^{-21}$ & $1.566 \times 10^{-22}$ \\
       90 & - & $3.377 \times 10^{-21}$ & - & - & - \\
      100 & $1.143 \times 10^{-20}$ & $2.486 \times 10^{-21}$ & $6.994 \times 10^{-22}$ & $1.374 \times 10^{-21}$ & $2.033 \times 10^{-22}$ \\
      150 & $2.343 \times 10^{-20}$ & $1.400 \times 10^{-21}$ & $7.170 \times 10^{-22}$ & $1.055 \times 10^{-21}$ & $1.678 \times 10^{-22}$ \\
      200 & $1.873 \times 10^{-20}$ & $1.478 \times 10^{-21}$ & $6.067 \times 10^{-22}$ & $8.486 \times 10^{-22}$ & $3.316 \times 10^{-22}$ \\
      300 & $2.873 \times 10^{-21}$ & $1.241 \times 10^{-21}$ & $3.878 \times 10^{-22}$ & $4.117 \times 10^{-22}$ & $5.604 \times 10^{-22}$ \\
\end{tabular}
\end{ruledtabular}
\caption{\label{tab:upper_limits}$\left<\sigma v\right>$ upper limits at the 90\% confidence level for all stacked dSphs. Notice the same physicality constraints as in TABLE~\ref{tab:energy_ranges}.}
}
\end{table*}

Signal PDFs are generated for each channel-mass combination. Note that these are source-independent, meaning the $\mathcal{S}_{k}\!\left(\psi_{i}\right)$ term in Eq.~(\ref{eq:LLH_stacked}) simplifies to $\mathcal{S}\!\left(\psi_{i}\right)$. The source-independence comes from the fact that we are only concerned about how far we expect a signal neutrino to be from a source, but this does not depend on where the source is in the sky; we also assume here that our angular reconstruction is uniform across the sky, which is justified given the 50\% containment of the median resolution is consistent among all source locations in this analysis. The detector acceptance for the exact location of each source is considered later when converting the number of signal events into a flux. For each neutrino in our Monte Carlo (MC) sample that passes the energy range cut for a given channel-mass combination, we calculate the angular reconstruction error between the MC true direction and the reconstructed direction. This is then weighted by $\frac{dN}{dE_{\nu}}\Big|_{E_{\text{MC}}}$ (that is, the value of the corresponding WIMP annihilation spectrum at the MC neutrino true energy) because we expect a neutrino to be more signal-like at larger values of $\frac{dN}{dE_{\nu}}$. The normalized histogram of all the weighted angular separations comprises the signal PDF. A sample of our signal PDFs is shown in Fig.~\ref{fig:signal_pdfs}.

%\FloatBarrier

\section{Results} \label{sec:results}

\begin{figure*}[ht!]
\centering
\includegraphics[width=0.9\linewidth]{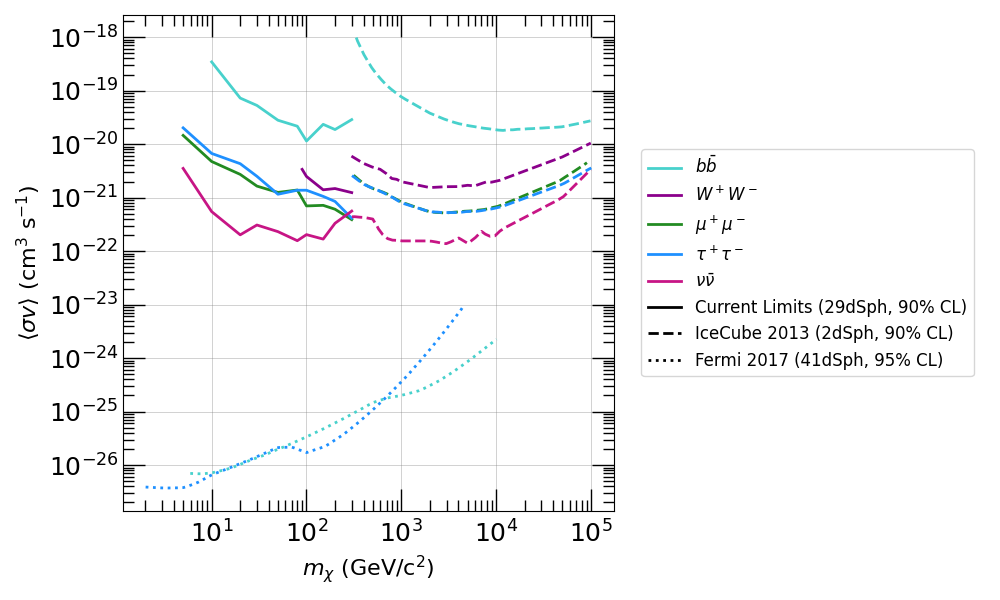}
\caption{\label{fig:upper_limits_comparison}Upper limits at 90\% confidence level (solid lines), with comparison to IceCube 2013~\cite{Aartsen2013} and Fermi 2017~\cite{Albert2017} results (dashed and dotted lines, respectively).}
\end{figure*}

For each channel-mass combination, we compute the best-fit number of signal events from WIMP annihilation by minimizing the negative of the LLH in Eq.~(\ref{eq:LLH_stacked}). We also compute the significance of each result by comparing the observed TS with a TS distribution obtained from background-only simulations. The highest TS is obtained for $20 \, \mathrm{GeV}$ WIMPs annihilating into a $\mu^{+}\mu^{-}$ pair, corresponding to a pre-trials significance of $2.3 \sigma$. We do not find any statistically-significant evidence of WIMP annihilation into neutrinos for any of the models that we study, and instead set upper limits at the $90\%$ confidence level. The upper limit for a given channel-mass combination is calculated by comparing the observed TS with a TS distribution obtained from an ensemble of simulations containing injected signal. The signal level for which the TS in $90\%$ of the simulations exceeds the observed TS is defined as the upper limit at the $90\%$ confidence level. This signal level is converted into a flux at the detector, which is then used in Eq.~(\ref{eq:wimp_annihilation}) to solve for $\left<\sigma v\right>$. These upper limits are listed in Table~\ref{tab:upper_limits}.

Our upper limits are shown in Fig.~\ref{fig:upper_limits_comparison} with comparisons to previous results from dwarf galaxies studies with IceCube~\cite{Aartsen2013} and Fermi~\cite{Albert2017}. Comparing to the previous IceCube results at the crossover mass of $300 \, \mathrm{GeV}$, we find that our limits for the $b\bar{b}$ channel have improved by nearly two orders of magnitude, and our limits for the $W^{+}W^{-}$, $\mu^{+}\mu^{-}$, and $\tau^{+}\tau^{-}$ channels have all improved by about half an order of magnitude. Our limits for the $\nu\bar{\nu}$ channel are comparable at this point. The primary drivers behind this improvement are the larger set of sources, additional neutrino data, the completion of the detector since the previous IceCube study, and improved reconstruction methods. This comparison is only made for WIMP masses of $\gtrsim 100 \, \mathrm{GeV}$ as this analysis is the first from IceCube to consider WIMP annihilation from dSphs down to the $\sim\!10 \, \mathrm{GeV}$ regime. The neutrino limits are less constraining at higher masses because the low-energy neutrino tail softens the neutrino spectra (see bottom panels of Fig.~\ref{fig:annihilation_spectra}). As a $\gamma$-ray detector, Fermi continues to set better limits in this low-mass regime due to higher effective area in this regime, less background contamination, and better angular resolution.

\begin{figure*}[htb]
\centering
\begin{subfigure}{0.45\textwidth}
\includegraphics[width=\linewidth]{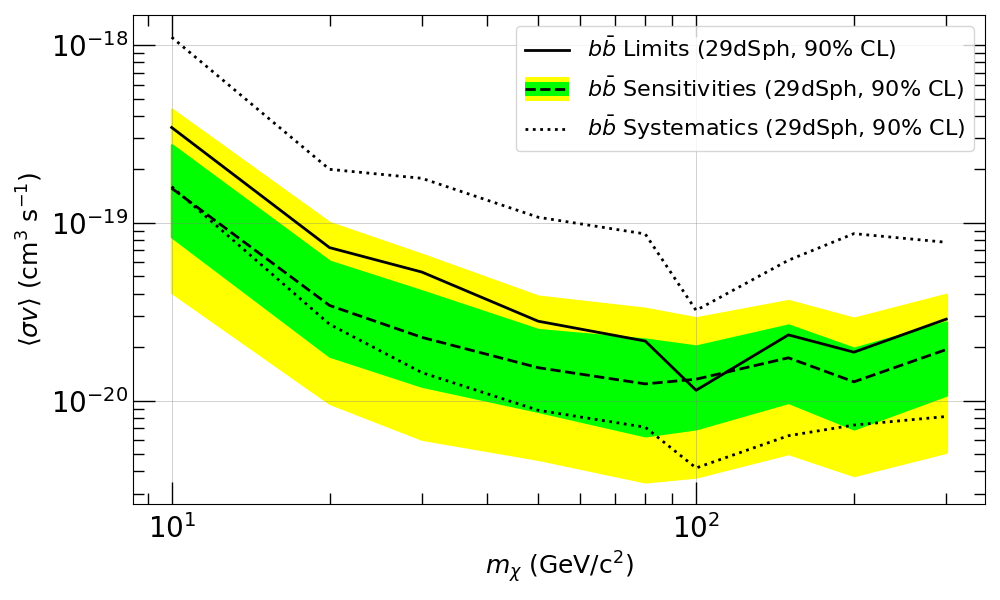}
\end{subfigure}
\begin{subfigure}{0.45\textwidth}
\includegraphics[width=\linewidth]{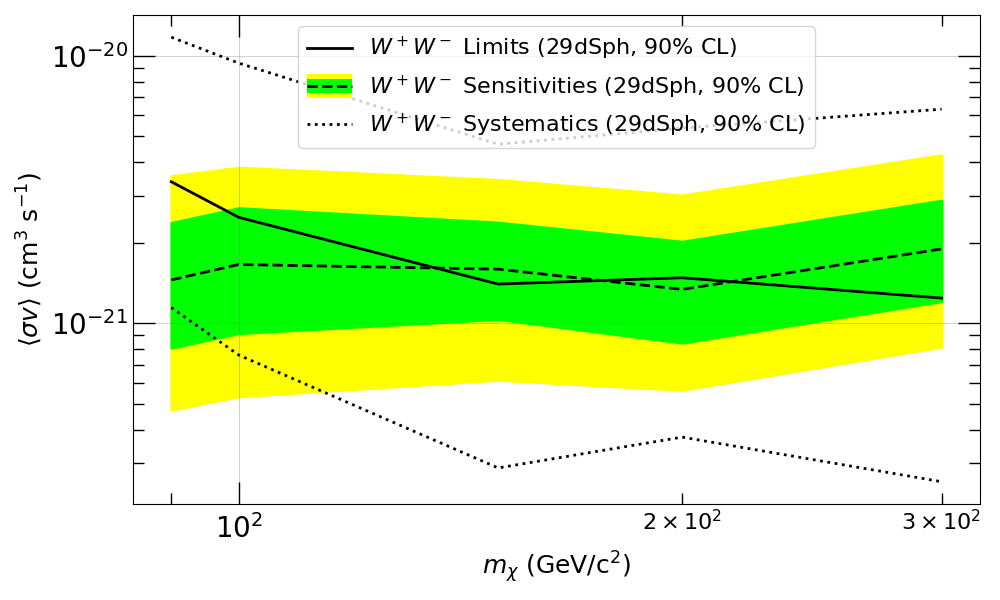}
\end{subfigure}
\begin{subfigure}{0.45\textwidth}
\includegraphics[width=\linewidth]{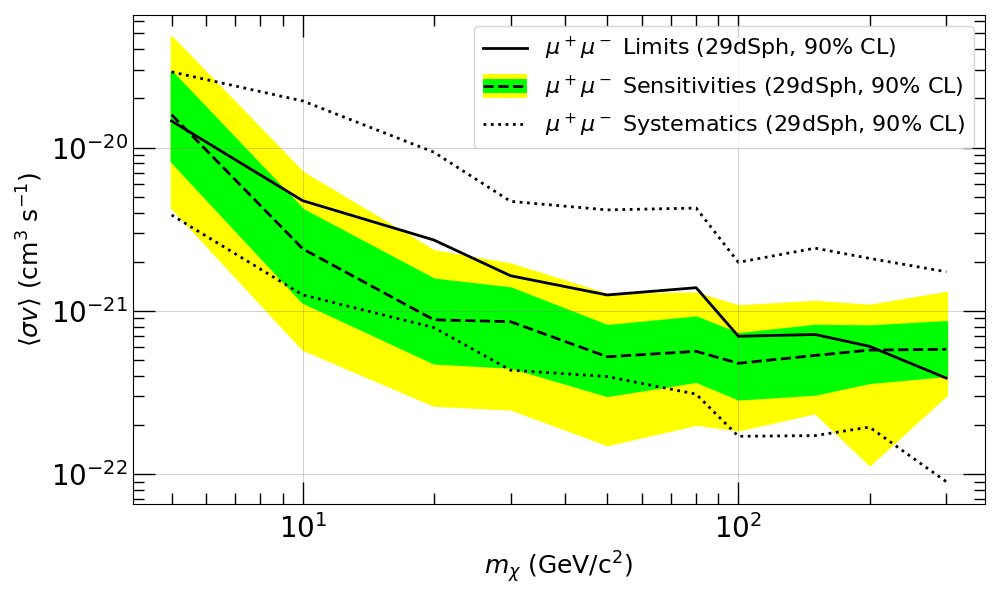}
\end{subfigure}
\begin{subfigure}{0.45\textwidth}
\includegraphics[width=\linewidth]{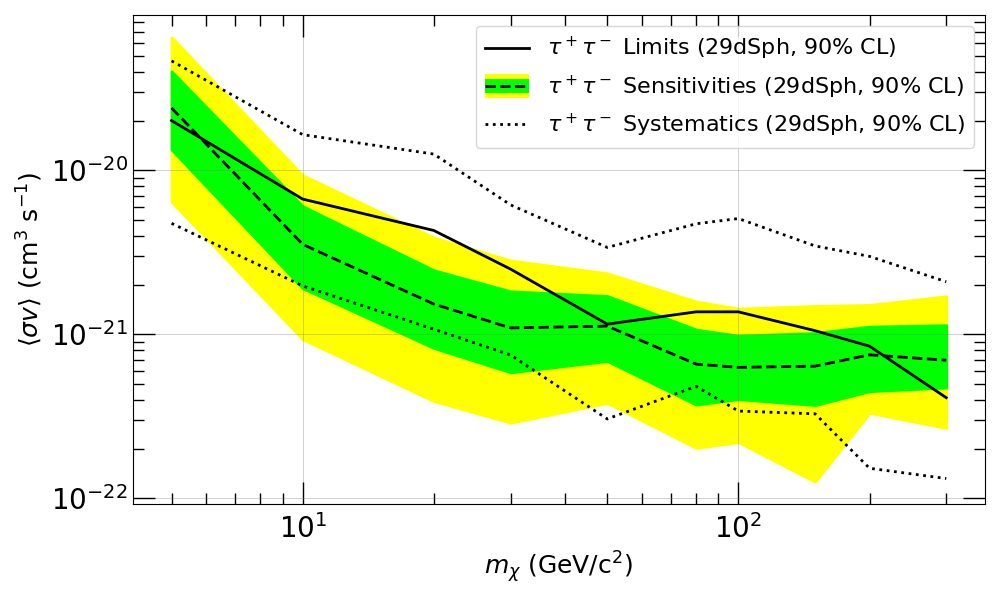}
\end{subfigure}
\begin{subfigure}{0.45\textwidth}
\includegraphics[width=\linewidth]{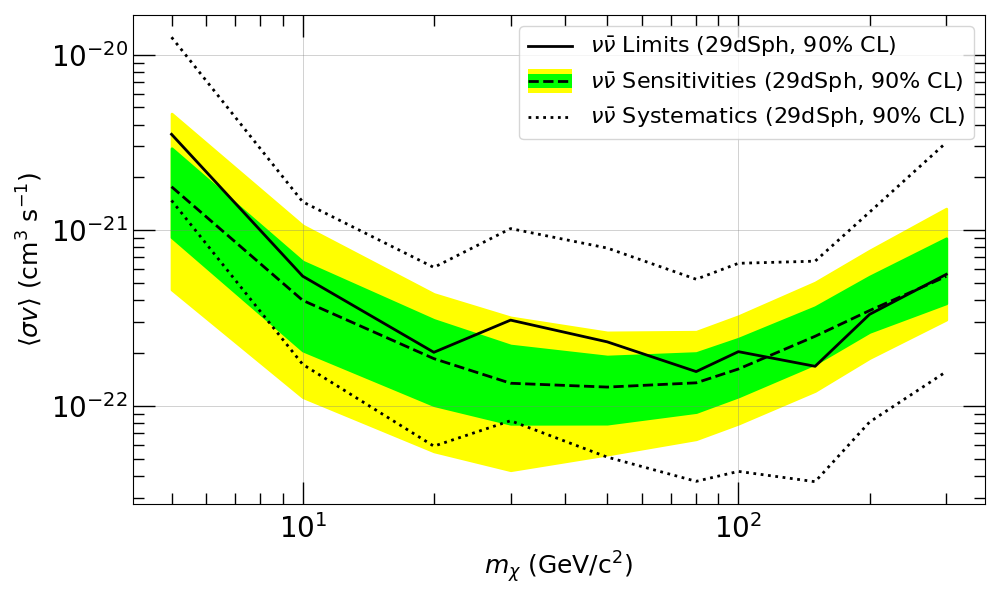}
\end{subfigure}
\caption{\label{fig:upper_limits_bands}Upper limits for individual channels (solid lines) with median sensitivities (dashed lines), $\pm 1\sigma$ and $2\sigma$ Brazil bands (shaded regions), and upper limits under systematic uncertainties (dotted lines).}
\end{figure*}

We also present the upper limits for each channel individually, as well as the corresponding median sensitivities and the "Brazil bands" (the $\pm 1\sigma$ and $\pm 2\sigma$ bands), both assuming a background-only expectation, in Fig.~\ref{fig:upper_limits_bands}. We find that our upper limits are consistent with our Brazil bands with the exception of three models: $\left(\mu^{+}\mu^{-}\right.$,$\left.20 \, \mathrm{GeV}\right)$, $\left(\mu^{+}\mu^{-}\right.$,$\left.80 \, \mathrm{GeV}\right)$, and $\left(\tau^{+}\tau^{-}\right.$,$\left.20 \, \mathrm{GeV}\right)$. These models all have stacked significances $>2\sigma$ $\left(\sigma = 2.382, 2.312, 2.270\right.$, respectively), which is why they lie outside the Brazil bands. A closer examination of the statistical fluctuations in these models is presented in Appendix~\ref{apx:post_unblinding_checks}.

\begin{figure*}[htbp]
\centering
\begin{subfigure}{\textwidth}
\includegraphics[width=\linewidth]{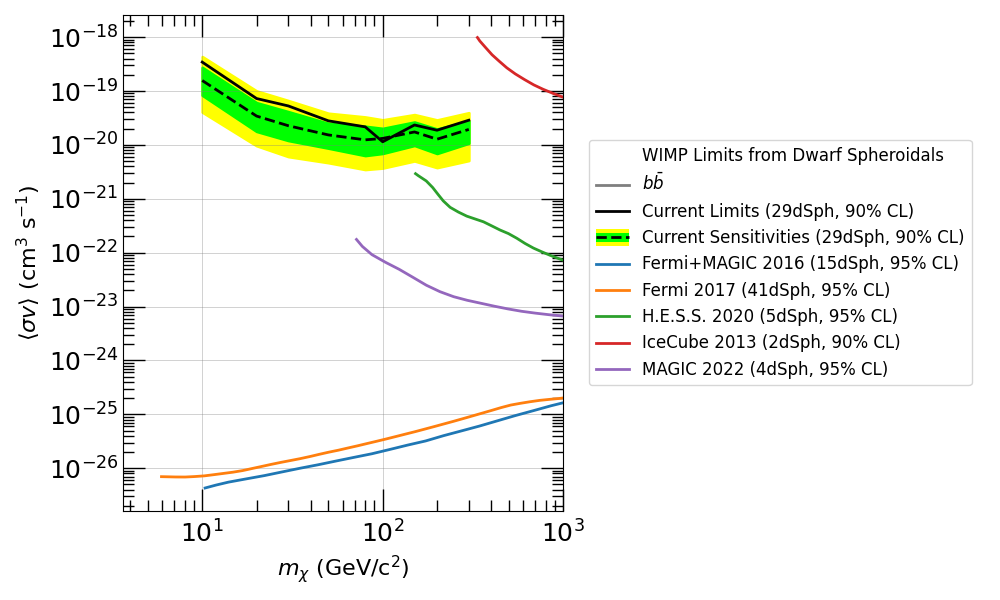}
\end{subfigure}
\begin{subfigure}{\textwidth}
\includegraphics[width=\linewidth]{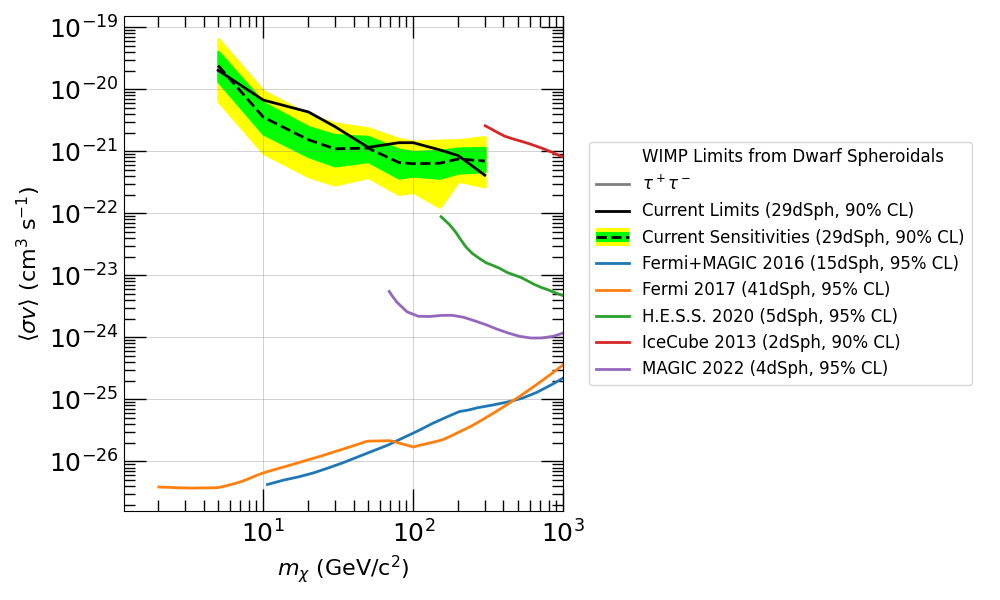}
\end{subfigure}
\caption{\label{fig:upper_limits_comparison_extended}Upper limits for $b\bar{b}$ (top) and $\tau^{+}\tau^{-}$ (bottom). Results from this work are shown in black in the same format as Fig.~\ref{fig:upper_limits_bands}. Comparisons are made to previous results from Fermi+MAGIC 2016~\cite{MAGIC2016}, Fermi 2017~\cite{Albert2017}, H.E.S.S 2020~\cite{Abdallah2020}, IceCube 2013~\cite{Aartsen2013}, and MAGIC 2022~\cite{Acciari2022}.}
\end{figure*}

Finally, we present a comparison of our results to a wider set of previous upper limits using dSphs as sources. These results are shown in Fig.~\ref{fig:upper_limits_comparison_extended}, with comparisons to previous results from Fermi+MAGIC 2016~\cite{MAGIC2016}, Fermi 2017~\cite{Albert2017}, H.E.S.S 2020~\cite{Abdallah2020}, IceCube 2013~\cite{Aartsen2013}, and MAGIC 2022~\cite{Acciari2022}. Again, the $\gamma$-ray detectors set stronger upper limits due to better sensitivity in this energy regime. However, these results highlight a large extension to lower masses among neutrino detectors, helping to explore the parameter space below $100 \, \mathrm{GeV}$.

%\FloatBarrier

\section{Systematic Uncertainties} \label{sec:systematic_uncertainties}

The largest systematic uncertainty in our analysis comes from the choice of J-factors used for the dSphs. Here, we present the corresponding limits assuming two systematic sets: $J\!\left(\theta_{\text{max}}\right)$ plus the corresponding upper uncertainty for each source, and $J\!\left(0.5^{\circ}\right)$ plus the lower uncertainty for each source; these uncertainties are listed with the J-factors in Table~\ref{tab:dwarfs}. These limits are presented in Fig.~\ref{fig:upper_limits_bands}. With slight variations due to statistical fluctuations, these systematic uncertainties are vertically offset from our upper limits by approximately half of an order of magnitude across all channels and masses that we consider.

%\FloatBarrier

\section{Conclusions} \label{sec:conclusions}

We present an analysis searching for signals of neutrinos produced via WIMP annihilation. We use \\ $\sim\!1.67 \times 10^{6}$ IceCube events with energies between $\sim\!10^{0.5}-10^{2.5} \, \mathrm{GeV}$. We test 5 different annihilation channels for WIMPs with masses between $5-300 \, \mathrm{GeV}$. We consider 29 different dSphs as sources of WIMP annihilation. We do not find any statistically significant evidence of WIMP annihilation into neutrinos, and instead set upper limits on the WIMP annihilation cross section $\left<\sigma v\right>$ at the 90\% confidence level. These are also IceCube's first limits below a WIMP mass of $300 \, \mathrm{GeV}$ from dSphs. While these results leveraged the DeepCore subarray, the IceCube Upgrade~\cite{Aartsen2019c,Abbasi2024}, a detector upgrade improving effective area and energy resolution down to $\mathcal{O}\!\left(1 \, \mathrm{GeV}\right)$, is currently being deployed. These improvements, along with discoveries of new dSphs and better characterizations of the J-factors of dSphs with surveys like LSST~\cite{He2015,Ando2019}, IceCube can probe additional DM parameter space and set stronger limits in the coming years.

\begin{acknowledgements}
The IceCube Collaboration acknowledges significant contributions to this manuscript from Brandon Pries. The authors gratefully acknowledge the support from the following agencies and institutions:
USA {\textendash} U.S. National Science Foundation-Office of Polar Programs,
U.S. National Science Foundation-Physics Division,
U.S. National Science Foundation-EPSCoR,
U.S. National Science Foundation-Office of Advanced Cyberinfrastructure,
Wisconsin Alumni Research Foundation,
Center for High Throughput Computing (CHTC) at the University of Wisconsin{\textendash}Madison,
Open Science Grid (OSG),
Partnership to Advance Throughput Computing (PATh),
Advanced Cyberinfrastructure Coordination Ecosystem: Services {\&} Support (ACCESS),
Frontera and Ranch computing project at the Texas Advanced Computing Center,
U.S. Department of Energy-National Energy Research Scientific Computing Center,
Particle astrophysics research computing center at the University of Maryland,
Institute for Cyber-Enabled Research at Michigan State University,
Astroparticle physics computational facility at Marquette University,
NVIDIA Corporation,
and Google Cloud Platform;
Belgium {\textendash} Funds for Scientific Research (FRS-FNRS and FWO),
FWO Odysseus and Big Science programmes,
and Belgian Federal Science Policy Office (Belspo);
Germany {\textendash} Bundesministerium f{\"u}r Bildung und Forschung (BMBF),
Deutsche Forschungsgemeinschaft (DFG),
Helmholtz Alliance for Astroparticle Physics (HAP),
Initiative and Networking Fund of the Helmholtz Association,
Deutsches Elektronen Synchrotron (DESY),
and High Performance Computing cluster of the RWTH Aachen;
Sweden {\textendash} Swedish Research Council,
Swedish Polar Research Secretariat,
Swedish National Infrastructure for Computing (SNIC),
and Knut and Alice Wallenberg Foundation;
European Union {\textendash} EGI Advanced Computing for research;
Australia {\textendash} Australian Research Council;
Canada {\textendash} Natural Sciences and Engineering Research Council of Canada,
Calcul Qu{\'e}bec, Compute Ontario, Canada Foundation for Innovation, WestGrid, and Digital Research Alliance of Canada;
Denmark {\textendash} Villum Fonden, Carlsberg Foundation, and European Commission;
New Zealand {\textendash} Marsden Fund;
Japan {\textendash} Japan Society for Promotion of Science (JSPS)
and Institute for Global Prominent Research (IGPR) of Chiba University;
Korea {\textendash} National Research Foundation of Korea (NRF);
Switzerland {\textendash} Swiss National Science Foundation (SNSF).
\end{acknowledgements}

\FloatBarrier

\clearpage

\appendix
\section{Probability Distribution Functions and Dataset} \label{apx:pdfs}

\begin{figure*}[ht!]
\centering
\begin{subfigure}{0.3\textwidth}
\includegraphics[width=\linewidth,height=4cm]{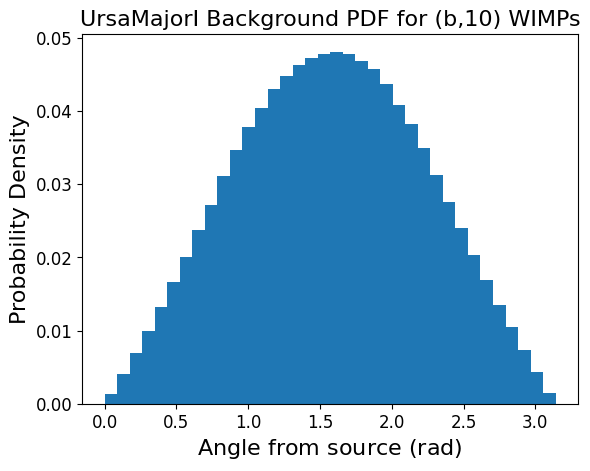}
\end{subfigure}
\begin{subfigure}{0.3\textwidth}
\includegraphics[width=\linewidth,height=4cm]{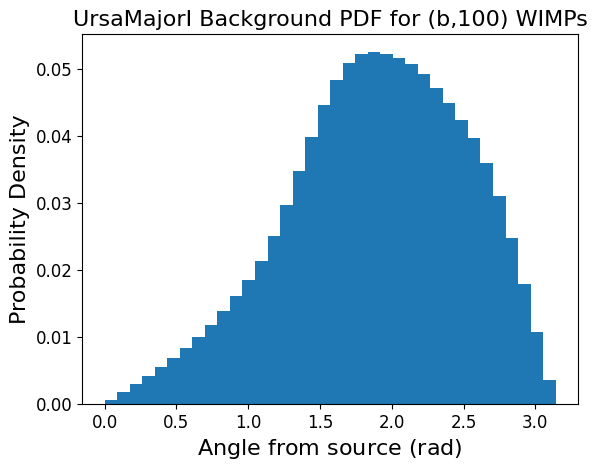}
\end{subfigure}
\begin{subfigure}{0.3\textwidth}
\includegraphics[width=\linewidth,height=4cm]{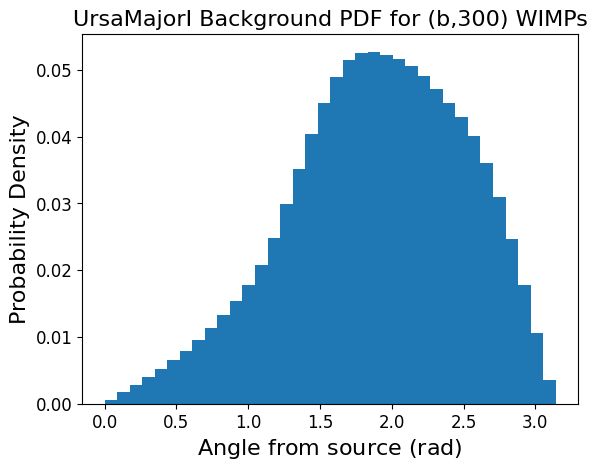}
\end{subfigure}
\begin{subfigure}{0.3\textwidth}
\includegraphics[width=\linewidth,height=4cm]{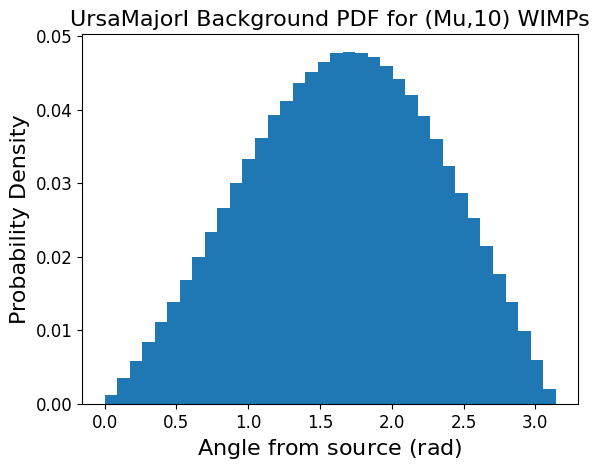}
\end{subfigure}
\begin{subfigure}{0.3\textwidth}
\includegraphics[width=\linewidth,height=4cm]{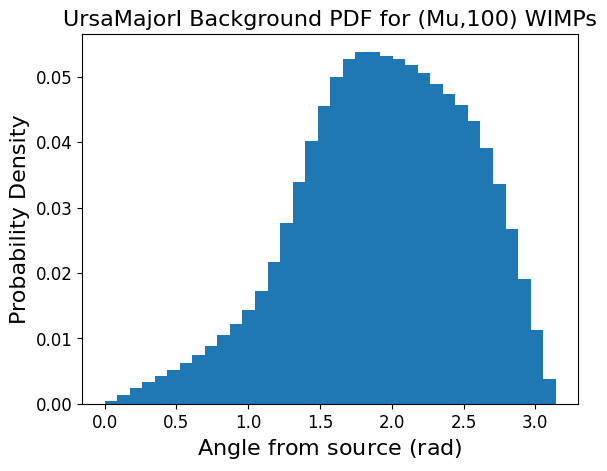}
\end{subfigure}
\begin{subfigure}{0.3\textwidth}
\includegraphics[width=\linewidth,height=4cm]{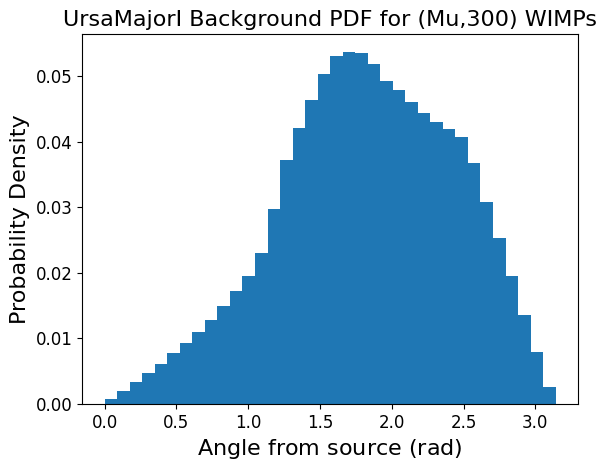}
\end{subfigure}
\begin{subfigure}{0.3\textwidth}
\includegraphics[width=\linewidth,height=4cm]{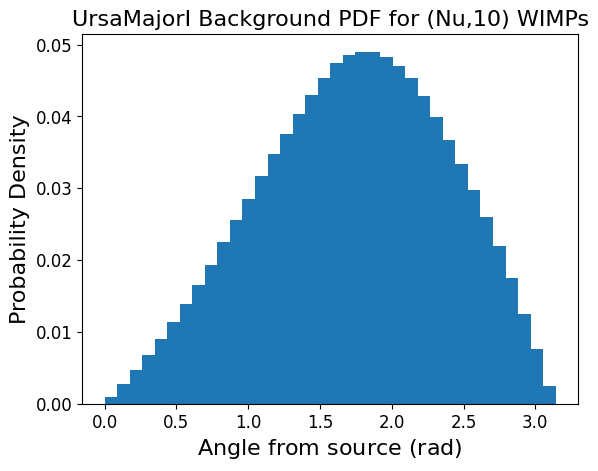}
\end{subfigure}
\begin{subfigure}{0.3\textwidth}
\includegraphics[width=\linewidth,height=4cm]{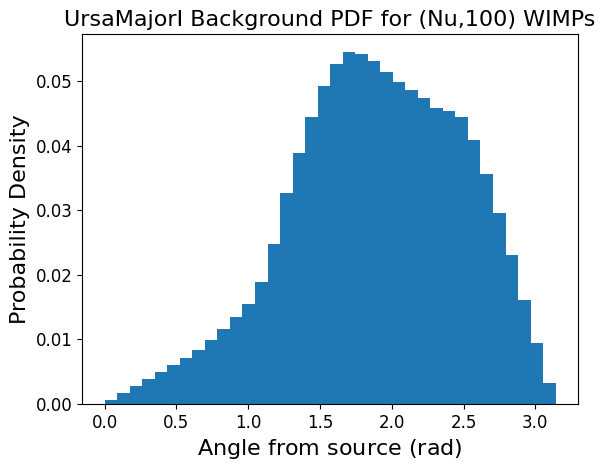}
\end{subfigure}
\begin{subfigure}{0.3\textwidth}
\includegraphics[width=\linewidth,height=4cm]{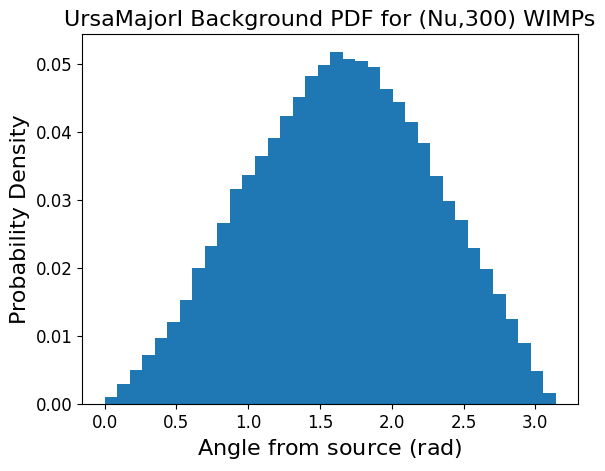}
\end{subfigure}
\caption{\label{fig:background_pdfs_UrsaMajorI}Background PDFs for selected channels and masses for Ursa Major I ($\delta \approx 50\degree$). Background PDFs for Ursa Major I are more peaked in the center than those for Leo IV (see Fig.~\ref{fig:background_pdfs_LeoIV}). At higher masses, the background PDFs shift to larger angles, which is a byproduct of Ursa Major I's northern declination and the DRAGON declination distribution being biased towards the southern sky.}
\end{figure*}

\begin{figure*}[ht!]
\centering
\begin{subfigure}{0.3\textwidth}
\includegraphics[width=\linewidth,height=4cm]{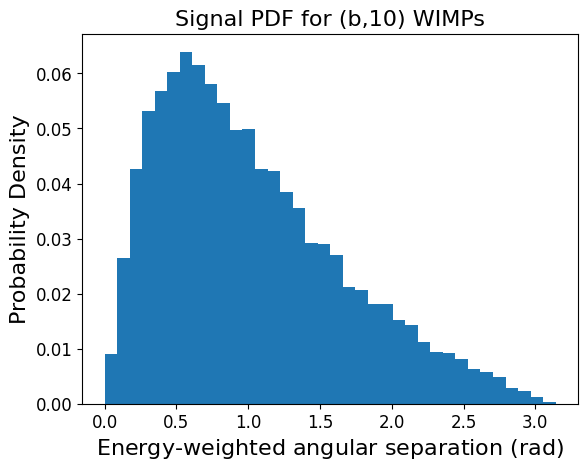}
\end{subfigure}
\begin{subfigure}{0.3\textwidth}
\includegraphics[width=\linewidth,height=4cm]{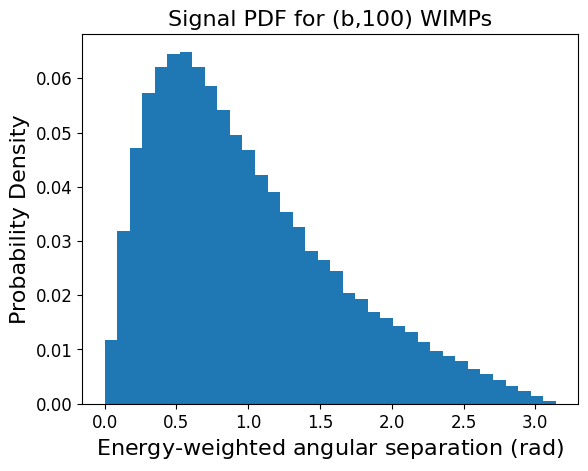}
\end{subfigure}
\begin{subfigure}{0.3\textwidth}
\includegraphics[width=\linewidth,height=4cm]{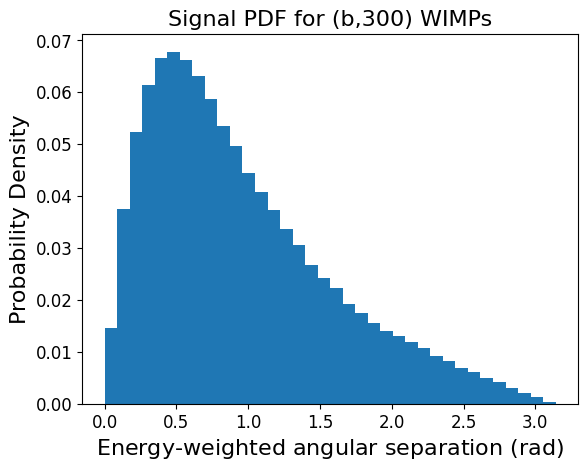}
\end{subfigure}
\begin{subfigure}{0.3\textwidth}
\includegraphics[width=\linewidth,height=4cm]{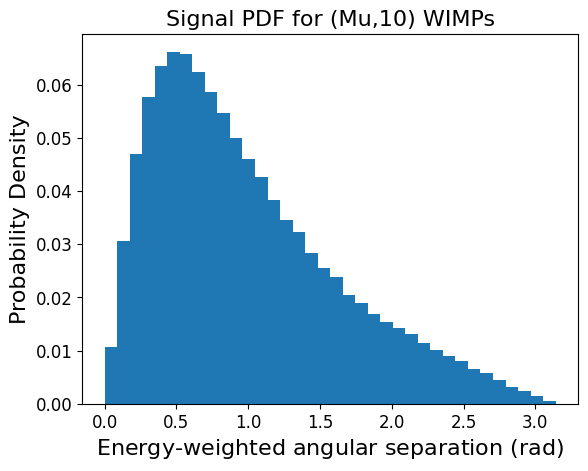}
\end{subfigure}
\begin{subfigure}{0.3\textwidth}
\includegraphics[width=\linewidth,height=4cm]{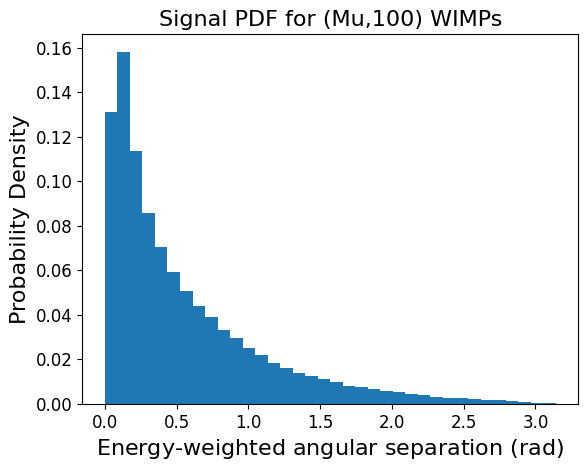}
\end{subfigure}
\begin{subfigure}{0.3\textwidth}
\includegraphics[width=\linewidth,height=4cm]{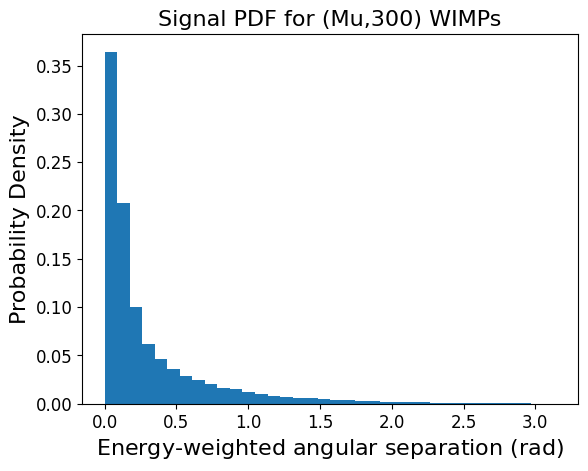}
\end{subfigure}
\begin{subfigure}{0.3\textwidth}
\includegraphics[width=\linewidth,height=4cm]{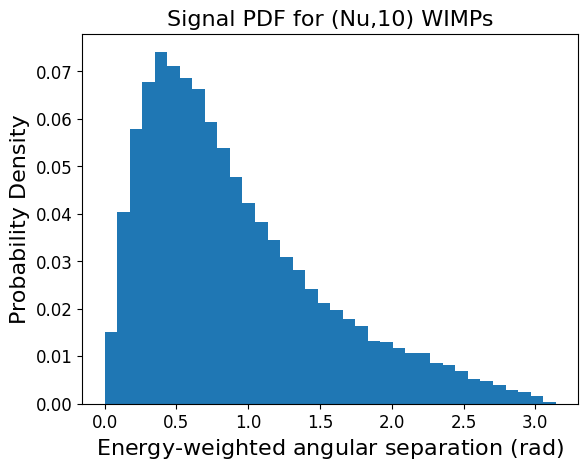}
\end{subfigure}
\begin{subfigure}{0.3\textwidth}
\includegraphics[width=\linewidth,height=4cm]{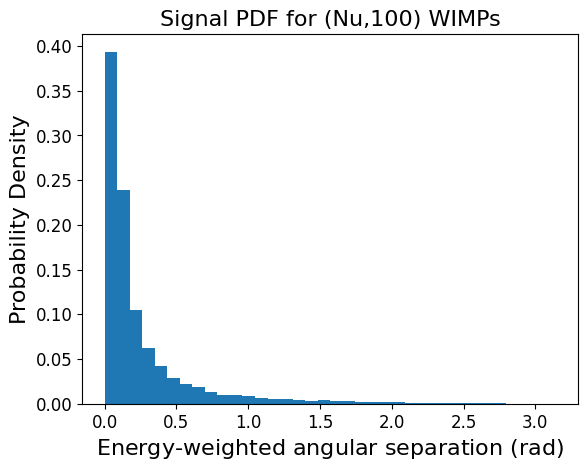}
\end{subfigure}
\begin{subfigure}{0.3\textwidth}
\includegraphics[width=\linewidth,height=4cm]{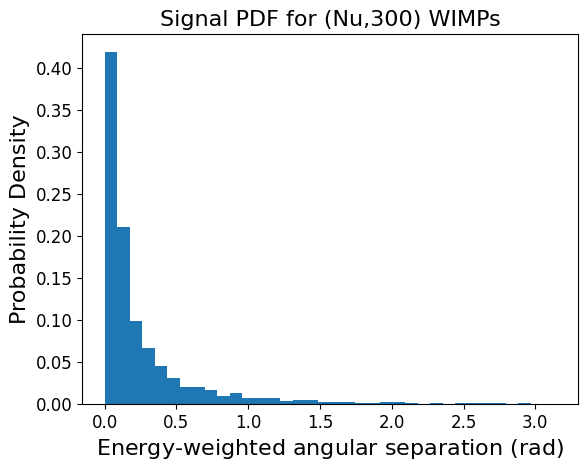}
\end{subfigure}
\caption{\label{fig:signal_pdfs}Signal PDFs for selected channels and masses. At low WIMP masses, signal PDFs have long tails that extend to high angular separations. For softer channels like $b\bar{b}$, the PDF smooths out at higher masses. This is because the energy range used to construct the PDF widens, increasing the number of neutrinos used to build in the PDF. For harder channels like $\mu^{+}\mu^{-}$ and $\nu\bar{\nu}$, the PDFs become more sharply peaked at low angular separations at higher masses. This is because the energy ranges shift upwards, excluding low-energy events with poorer angular reconstructions.}
\end{figure*}

\begin{figure*}[ht!]
\centering
\includegraphics[width=0.8\textwidth]{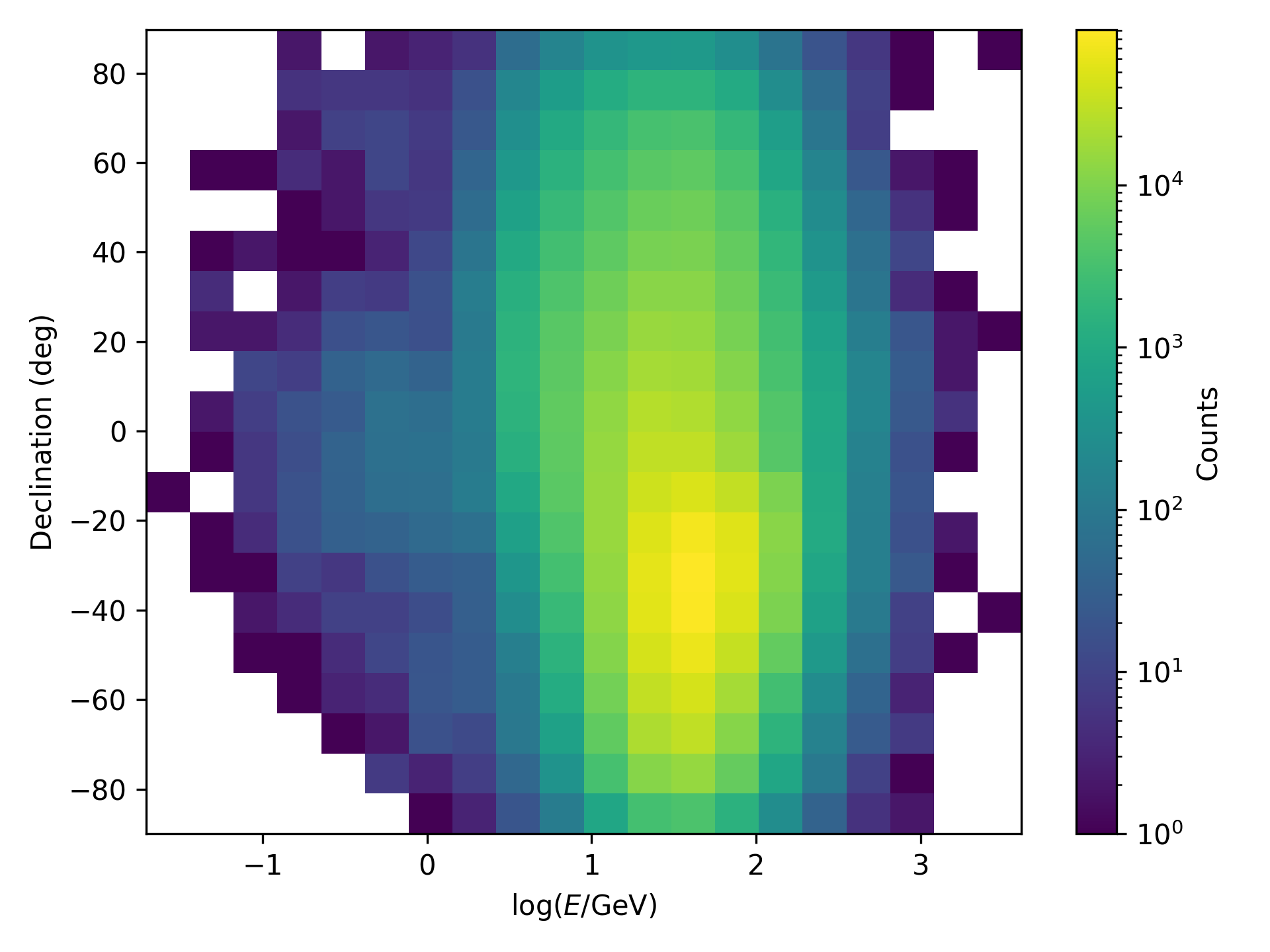}
\caption{\label{fig:dec_logE}2D histogram of neutrino energy and neutrino declination. The distribution highlights a strong bias towards the southern sky for neutrinos between $\sim\!10-100 \, \mathrm{GeV}$. This distribution, in addition to the energy range cuts shown in Table~\ref{tab:energy_ranges}, determine the shapes of the PDFs.}
\end{figure*}

Here we present samples of our background and signal PDFs, not shown in the main body of the paper for brevity. Our background PDFs for Leo IV and Ursa Major I are shown in Fig.~\ref{fig:background_pdfs_LeoIV} and Fig.~\ref{fig:background_pdfs_UrsaMajorI}, respectively. Leo IV was chosen for its proximity to the ecliptic $\left(\delta = -0.53^{\circ}\right)$, and Ursa Major I was selected to highlight a source roughly halfway between the ecliptic and the north celestial pole $\left(\delta = 51.92^\circ\right)$. Our signal PDFs are shown in Fig.~\ref{fig:signal_pdfs}. We also present the declination and energy dependence of the neutrinos in our dataset in Fig.~\ref{fig:dec_logE}.

\onecolumngrid

\vspace{1in}

\begin{center}
{\captionsetup{type=figure}
\captionsetup[sub]{labelformat=empty,labelsep=none,list=false}
\subcaptionbox{}{%
    \includegraphics[width=0.3\textwidth,height=4cm]{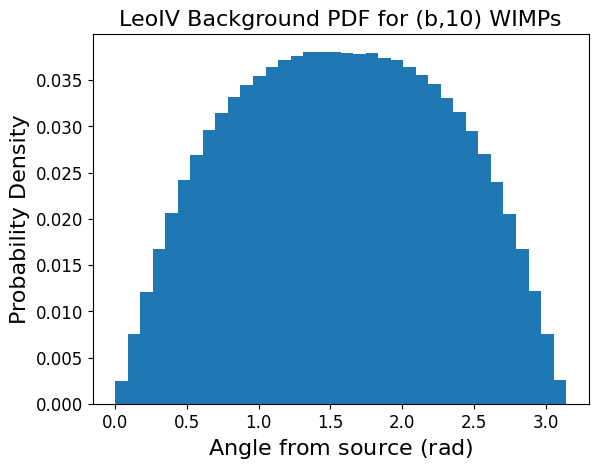}
    \hfill
    \includegraphics[width=0.3\textwidth,height=4cm]{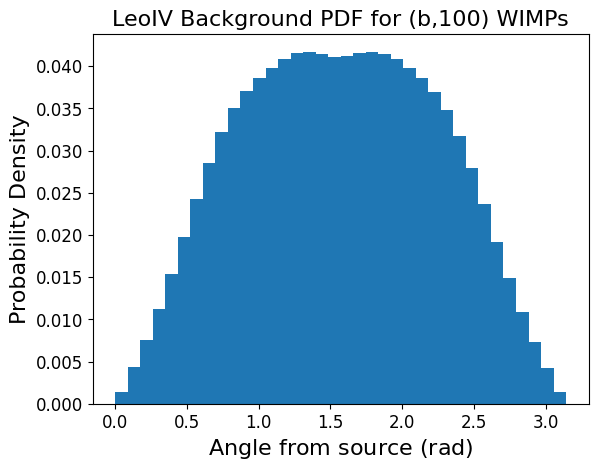}
    \hfill
    \includegraphics[width=0.3\textwidth,height=4cm]{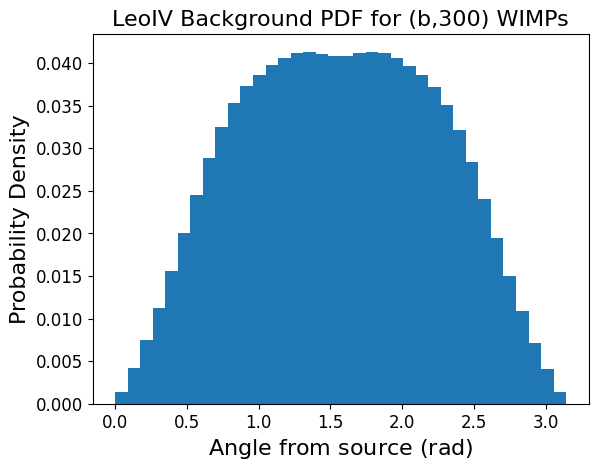}
}
\subcaptionbox{}{%
    \includegraphics[width=0.3\textwidth,height=4cm]{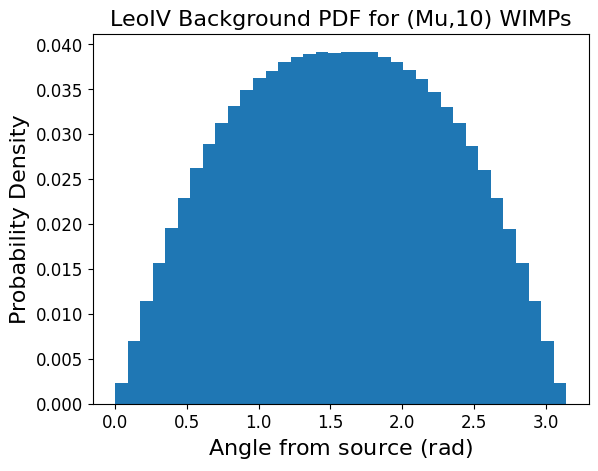}
    \hfill
    \includegraphics[width=0.3\textwidth,height=4cm]{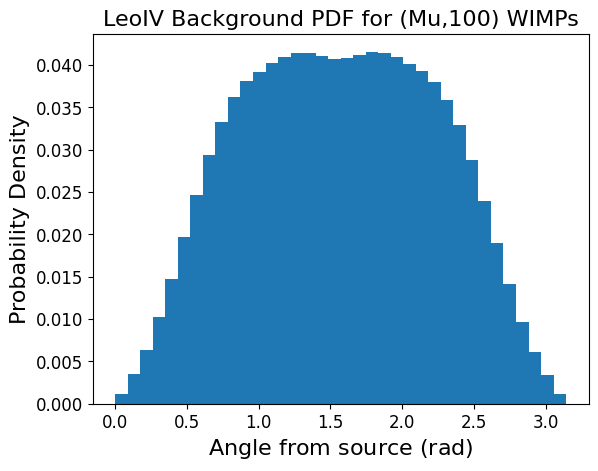}
    \hfill
    \includegraphics[width=0.3\textwidth,height=4cm]{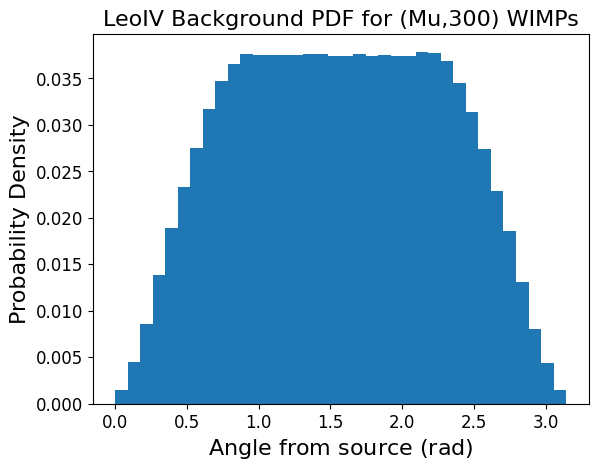}
}
\subcaptionbox{}{%
    \includegraphics[width=0.3\textwidth,height=4cm]{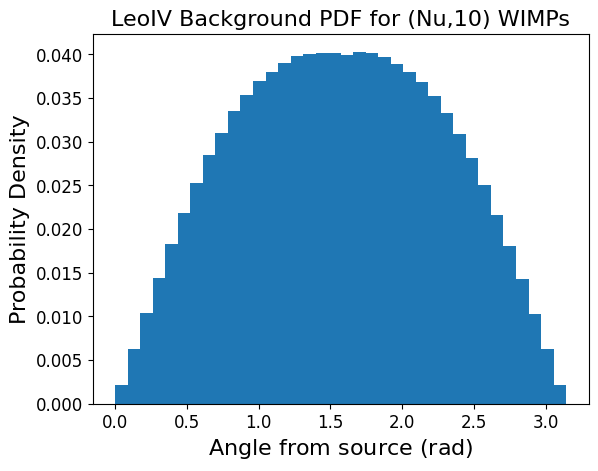}
    \hfill
    \includegraphics[width=0.3\textwidth,height=4cm]{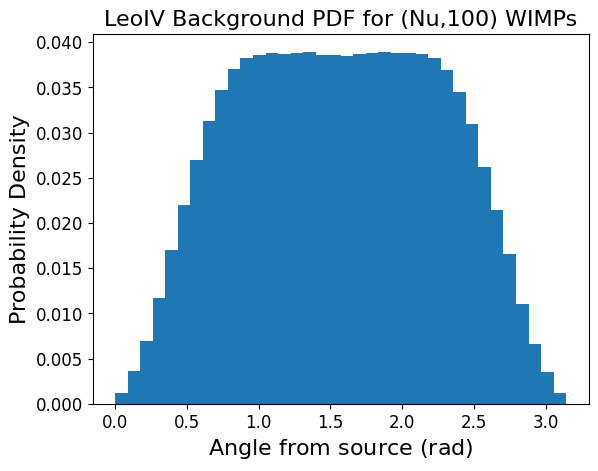}
    \hfill
    \includegraphics[width=0.3\textwidth,height=4cm]{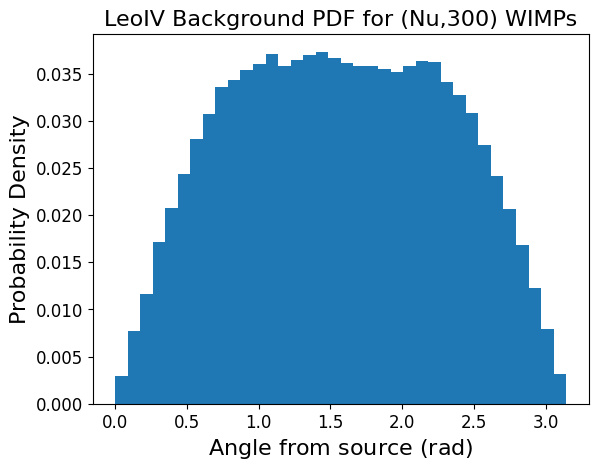}
}
}
\captionof{figure}{Background PDFs for selected channels and masses for Leo IV ($\delta \approx 0\degree$). Background PDFs show a large central peak at low masses. This shifts to a bimodal distribution with a small dip at a separation of $\theta \approx 90^{\circ}$, which is due to declination and phase space effects: for a source near the ecliptic, this region coincides with the celestial poles, which contain fewer events.} \label{fig:background_pdfs_LeoIV}
\end{center}
\twocolumngrid

\clearpage

\FloatBarrier

\section{Additional Cross-Checks} \label{apx:post_unblinding_checks}

\begin{figure*}[ht!]
\centering
\begin{subfigure}{0.3\textwidth}
\includegraphics[width=\textwidth,height=4cm]{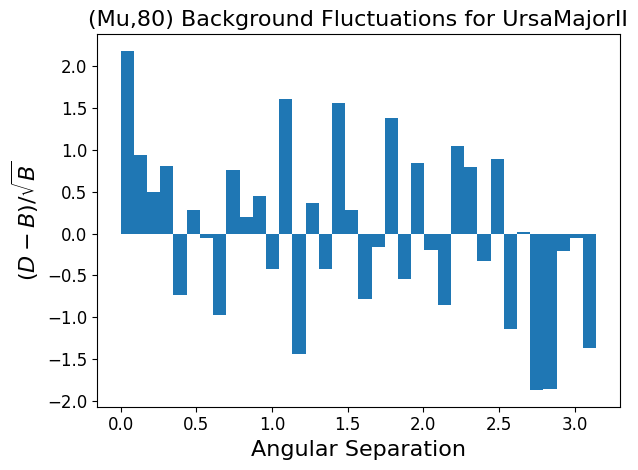}
\end{subfigure}
\begin{subfigure}{0.3\textwidth}
\includegraphics[width=\textwidth,height=4cm]{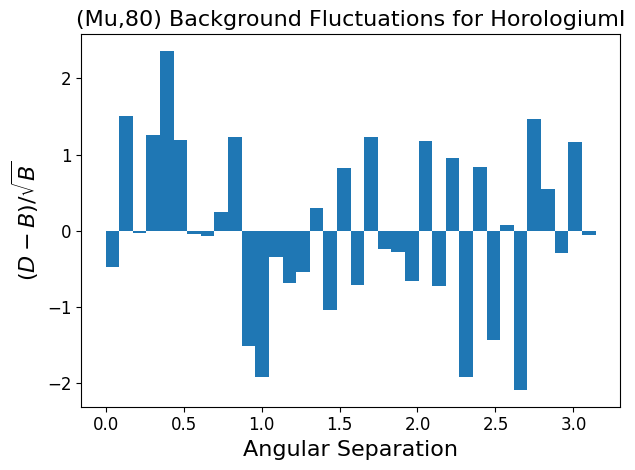}
\end{subfigure}
\begin{subfigure}{0.3\textwidth}
\includegraphics[width=\textwidth,height=4cm]{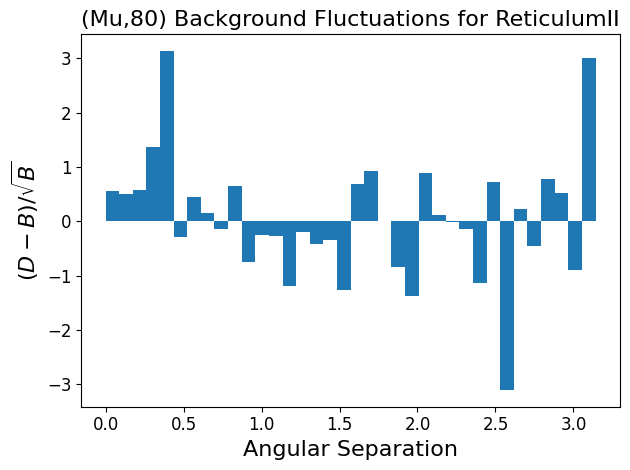}
\end{subfigure}
\begin{subfigure}{0.3\textwidth}
\includegraphics[width=\textwidth,height=4cm]{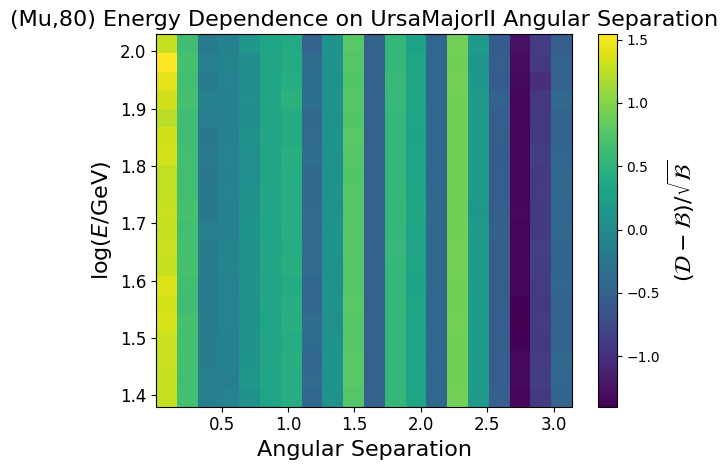}
\end{subfigure}
\begin{subfigure}{0.3\textwidth}
\includegraphics[width=\textwidth,height=4cm]{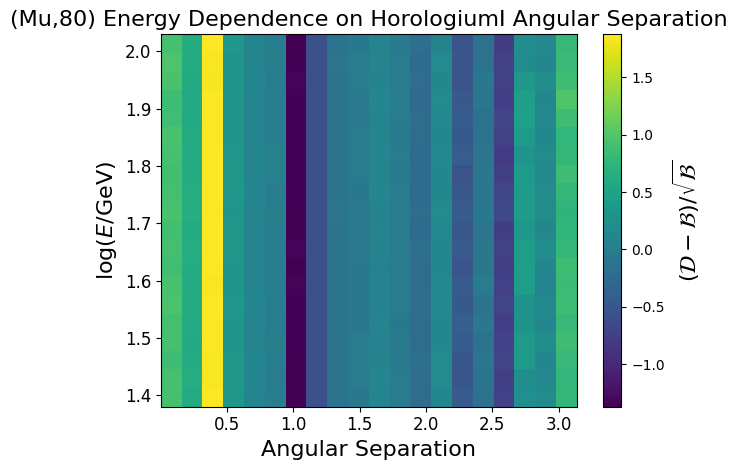}
\end{subfigure}
\begin{subfigure}{0.3\textwidth}
\includegraphics[width=\textwidth,height=4cm]{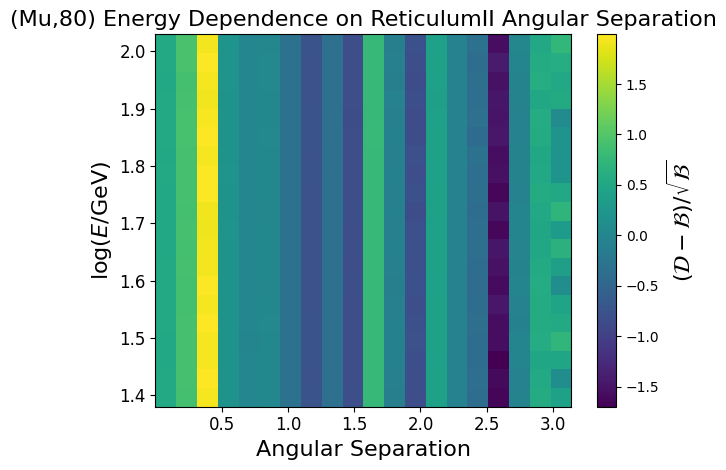}
\end{subfigure}
\caption{\label{fig:unblinding_Mu_80}Same as Fig.~\ref{fig:unblinding_Mu_20} for the $\mu^{+}\mu^{-}$ annihilation channel at $80 \, \mathrm{GeV}$.}
\end{figure*}

\begin{figure*}[ht!]
\centering
\begin{subfigure}{0.3\textwidth}
\includegraphics[width=\textwidth,height=4cm]{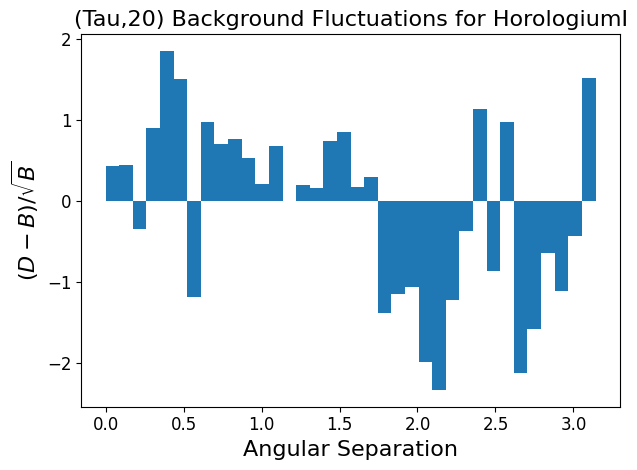}
\end{subfigure}
\begin{subfigure}{0.3\textwidth}
\includegraphics[width=\textwidth,height=4cm]{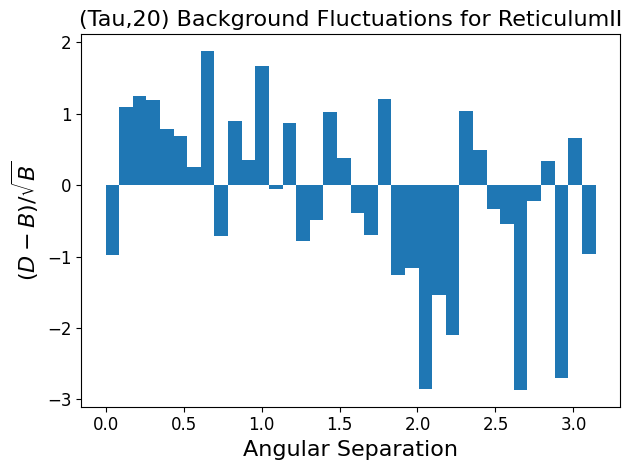}
\end{subfigure}
\begin{subfigure}{0.3\textwidth}
\includegraphics[width=\textwidth,height=4cm]{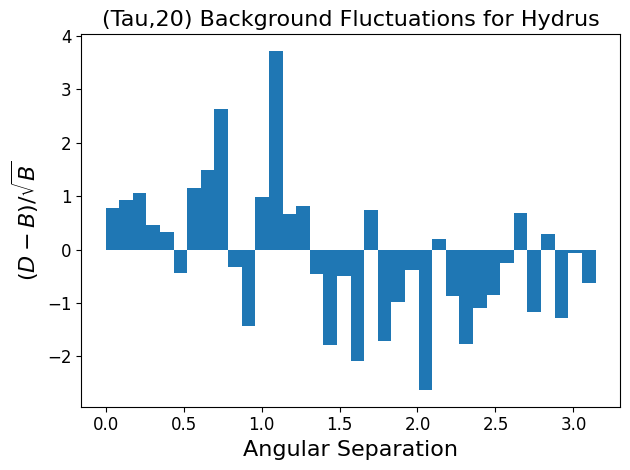}
\end{subfigure}
\begin{subfigure}{0.3\textwidth}
\includegraphics[width=\textwidth,height=4cm]{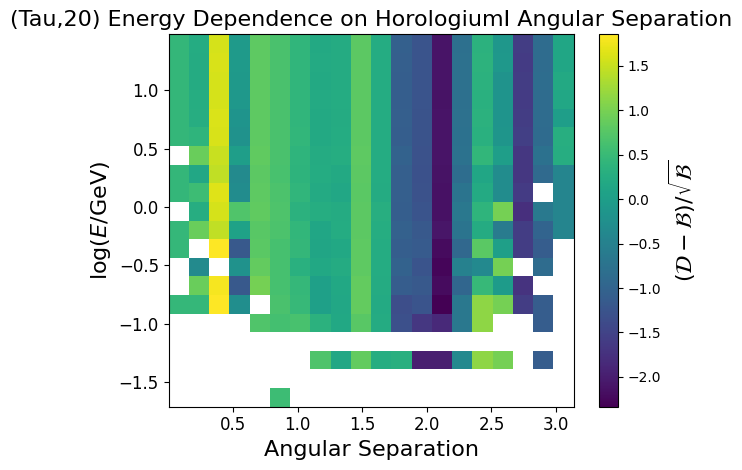}
\end{subfigure}
\begin{subfigure}{0.3\textwidth}
\includegraphics[width=\textwidth,height=4cm]{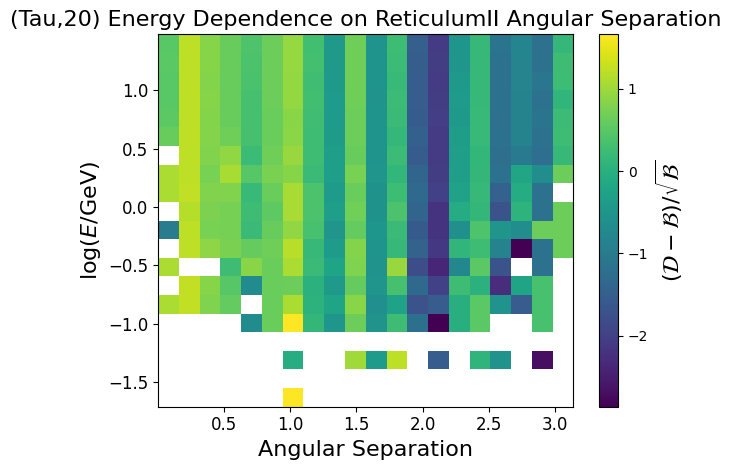}
\end{subfigure}
\begin{subfigure}{0.3\textwidth}
\includegraphics[width=\textwidth,height=4cm]{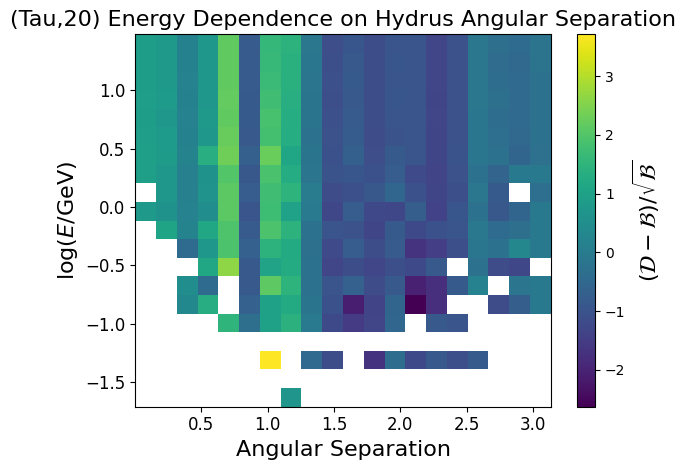}
\end{subfigure}
\caption{\label{fig:unblinding_Tau_20}Same as Fig.~\ref{fig:unblinding_Mu_20} for the $\tau^{+}\tau^{-}$ annihilation channel at $20 \, \mathrm{GeV}$.}
\end{figure*}

Here, we further inspect the data for channel-mass combinations that showed upwards fluctuations exceeding $2\sigma$. We show the data around the three dSphs contributing the highest number of signal events to our fit for each channel-mass combination; we note that these dSphs are typically considered "ultrafaint" dSphs in the literature and tend to have larger J-factor uncertainties that those considered to be "classical" dSphs such as Draco, Fornax, Sculptor, and Ursa Minor (see Table~\ref{tab:dwarfs}). As shown in figures \ref{fig:unblinding_Mu_20},\ref{fig:unblinding_Mu_80} and \ref{fig:unblinding_Tau_20}, positive statistical fluctuations in the vicinity of the sources contribute to the $\sim 2\sigma$ pre-trials significance values obtained for these models.

% \begin{figure*}[ht!]
% \centering
% \begin{subfigure}{0.3\textwidth}
% \includegraphics[width=\textwidth,height=4cm]{Images/sigmas_bkg_angsep_Mu_20_HorologiumI_bigfont.png}
% \end{subfigure}
% \begin{subfigure}{0.3\textwidth}
% \includegraphics[width=\textwidth,height=4cm]{Images/sigmas_bkg_angsep_Mu_20_ReticulumII_bigfont.png}
% \end{subfigure}
% \begin{subfigure}{0.3\textwidth}
% \includegraphics[width=\textwidth,height=4cm]{Images/sigmas_bkg_angsep_Mu_20_Hydrus_bigfont.png}
% \end{subfigure}
% \begin{subfigure}{0.3\textwidth}
% \includegraphics[width=\textwidth,height=4cm]{Images/DRAGON_logE_angsep_weighted_bkg_Mu_20_HorologiumI_bigfont.png}
% \end{subfigure}
% \begin{subfigure}{0.3\textwidth}
% \includegraphics[width=\textwidth,height=4cm]{Images/DRAGON_logE_angsep_weighted_bkg_Mu_20_ReticulumII_bigfont.png}
% \end{subfigure}
% \begin{subfigure}{0.3\textwidth}
% \includegraphics[width=\textwidth,height=4cm]{Images/DRAGON_logE_angsep_weighted_bkg_Mu_20_Hydrus_bigfont.png}
% \end{subfigure}
% \caption{\label{fig:unblinding_Mu_20}Cross-checks for the $\mu^{+}\mu^{-}$ annihilation channel at $20 \, \mathrm{GeV}$ highlighting source of fluctuations above background expectations. The upper row are pulls relative to background expectations binned in angular separation. The lower row is the top row projected onto the corresponding $\log(E/\mathrm{GeV})$ vs.\ angular separation histogram. Columns correspond to the three sources contributing the highest number of signal events to the analysis, ranked from left to right.}
% \end{figure*}

\onecolumngrid

\vspace{1in}

\begin{center}
{\captionsetup{type=figure}
\captionsetup[sub]{labelformat=empty,labelsep=none,list=false}
\subcaptionbox{}{%
    \includegraphics[width=0.3\textwidth,height=4cm]{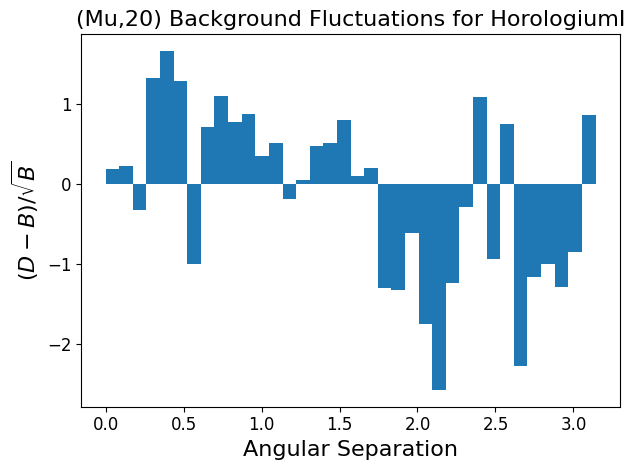}
    \hfill
    \includegraphics[width=0.3\textwidth,height=4cm]{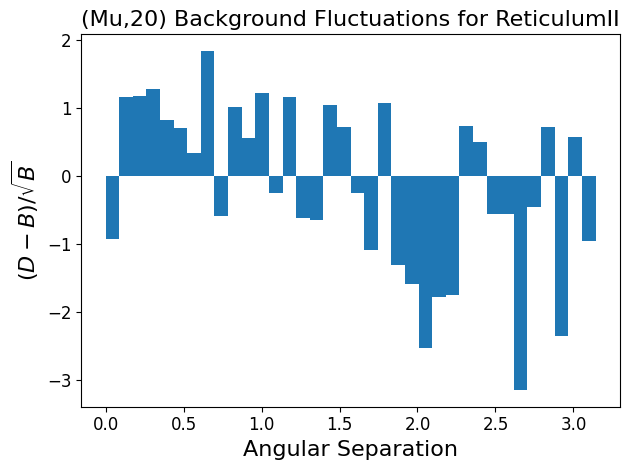}
    \hfill
    \includegraphics[width=0.3\textwidth,height=4cm]{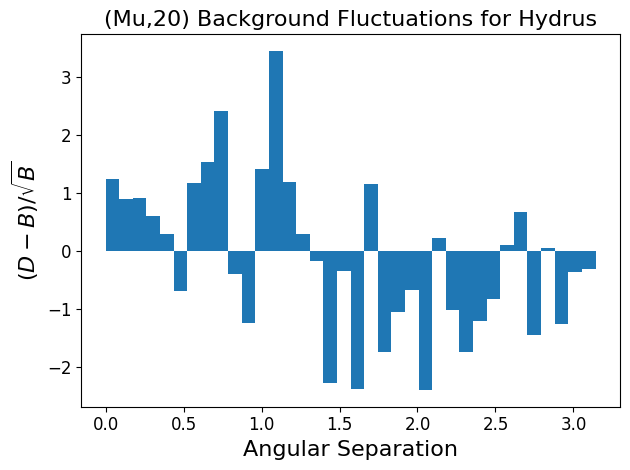}
}
\subcaptionbox{}{%
    \includegraphics[width=0.3\textwidth,height=4cm]{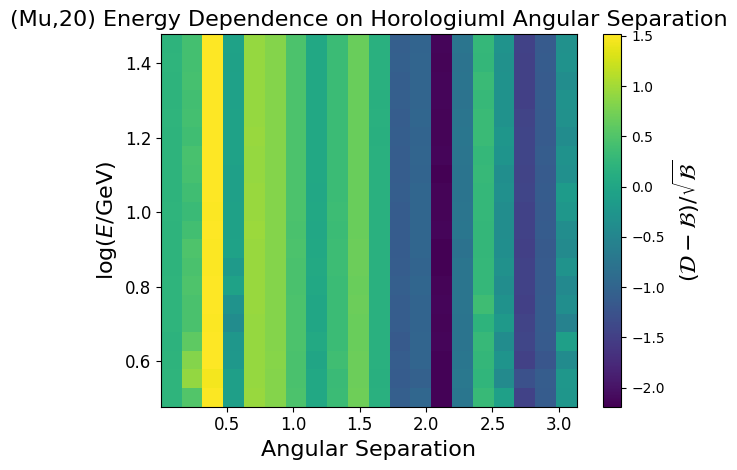}
    \hfill
    \includegraphics[width=0.3\textwidth,height=4cm]{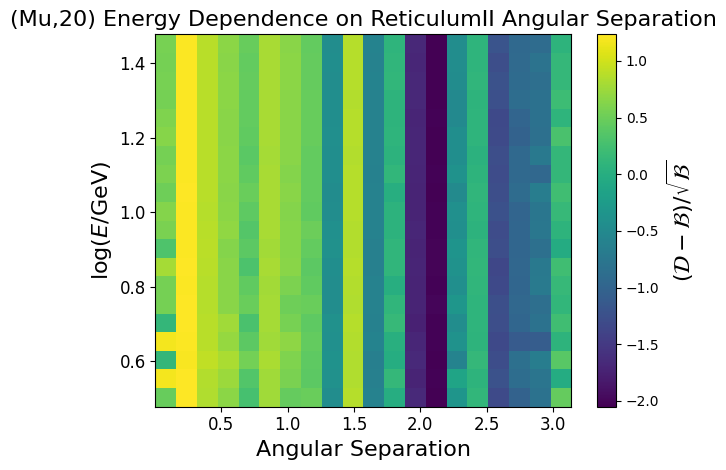}
    \hfill
    \includegraphics[width=0.3\textwidth,height=4cm]{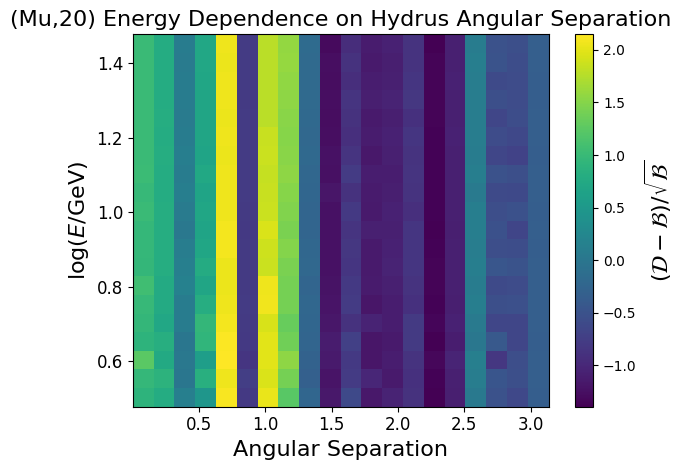}
}
}
\captionof{figure}{Cross-checks for the $\mu^{+}\mu^{-}$ annihilation channel at $20 \, \mathrm{GeV}$ highlighting source of fluctuations above background expectations. The upper row are pulls relative to background expectations binned in angular separation. The lower row is the top row projected onto the corresponding $\log(E/\mathrm{GeV})$ vs.\ angular separation histogram. Columns correspond to the three sources contributing the highest number of signal events to the analysis, ranked from left to right.} \label{fig:unblinding_Mu_20}
\end{center}
\twocolumngrid

\clearpage

\FloatBarrier

\clearpage

\bibliography{refs}

\end{document}